\documentclass[twocolumn,tighten]{aastex62}
\hypersetup{linkcolor=red,citecolor=blue,filecolor=cyan,urlcolor=magenta}
\usepackage{amsmath}
\usepackage{natbib}
\usepackage{color}
\usepackage{hyperref}
\usepackage{url}

\newcommand{\logg}{\mbox{log~{\it g}}}
\newcommand{\msun}{\mbox{$M_{\odot}$}}
\newcommand{\teff}{\mbox{$T_{\rm eff}$}}

\def\vector#1{\mbox{\boldmath $#1$}}
\newcommand{\AU}{\ensuremath{\,\mathrm{AU}}}
\newcommand{\pc}{\ensuremath{\,\mathrm{pc}}}
\newcommand{\kpc}{\ensuremath{\,\mathrm{kpc}}}
\newcommand{\Myr}{\ensuremath{\,\mathrm{Myr}}}
\newcommand{\Gyr}{\ensuremath{\,\mathrm{Gyr}}}
\newcommand{\kms}{\ensuremath{\,\mathrm{km\ s}^{-1}}}

\newcommand{\mas}{\ensuremath{\,\mathrm{mas}}}
\newcommand{\masyr}{\ensuremath{\,\mathrm{mas\ yr}^{-1}}}

\newcommand{\Kelvin}{\ensuremath{\,\mathrm{K}}}
\newcommand{\vlos}{v_{\ensuremath{\mathrm{los}}}}

\newcommand{\mualpha}{\mu_{\alpha*}}
\newcommand{\mudelta}{\mu_\delta}
\newcommand{\rperi}{r_\ensuremath{\mathrm{peri}}}

\newcommand{\Gaia}{\textit{Gaia}}

\newcommand{\eq}[1]{\begin{align}#1\end{align}}

\newcommand{\hvs}{\object[2MASS J09120652+0916216]{LAMOST-HVS1}}
\newcommand{\hvstwo}{\object[2MASS 16202076+3747399]{LAMOST-HVS2}}%[PG 1618+379]

\newcommand{\hvsfour}{\object[LAMOST-HVS4]{LAMOST-HVS4}}
\newcommand{\runawayI}{\object[HIP 114569]{HIP 114569}}
\newcommand{\runawayII}{\object[HIP 11809]{HIP 11809}}
\newcommand{\runawayIII}{\object[PG 1533+467]{PG 1533+467}}
\newcommand{\hd}{\object[HD 271791]{HD 271791}}

\shorttitle{Hyper-runaway star LAMOST-HVS1}
\shortauthors{Hattori et al.}
\begin{document}

\title{%
Origin of a massive hyper-runaway subgiant star LAMOST-HVS1 
-- implication from \Gaia\ and follow-up spectroscopy
}

\author{Kohei Hattori}
\affiliation{%
Department of Astronomy, University of Michigan,
1085 S.\ University Ave., Ann Arbor, MI 48109, USA}
\email{Email:\ khattori@umich.edu}

\author{Monica Valluri}
\affiliation{%
Department of Astronomy, University of Michigan,
1085 S.\ University Ave., Ann Arbor, MI 48109, USA}

\author{Norberto Castro}
\affiliation{%
Department of Astronomy, University of Michigan,
1085 S.\ University Ave., Ann Arbor, MI 48109, USA}
\affiliation{%
Leibniz-Institut f\"{u}r Astrophysik Potsdam (AIP), 
An der Sternwarte 16, 14482, Potsdam, Germany}

\author{Ian U.\ Roederer}
\affiliation{%
Department of Astronomy, University of Michigan,
1085 S.\ University Ave., Ann Arbor, MI 48109, USA}
\affiliation{%
Joint Institute for Nuclear Astrophysics -- Center for the
Evolution of the Elements (JINA-CEE), USA}

\author{Guillaume Mahler}
\affiliation{%
Department of Astronomy, University of Michigan,
1085 S.\ University Ave., Ann Arbor, MI 48109, USA}

\author{Gourav~Khullar}
\affiliation{%
Department of Astronomy and Astrophysics, University of
Chicago, 5640 South Ellis Avenue, Chicago, IL 60637, USA}
\affiliation{%
Kavli Institute of Cosmological Physics, University of
Chicago, 5640 South Ellis Avenue, Chicago, IL 60637, USA}

\begin{abstract}

We report that \hvs\ 
is a massive hyper-runaway subgiant star with mass of $8.3~\msun$ 
and super-Solar metallicity, 
ejected from the inner stellar disk of the Milky Way $\sim 33 \Myr$ ago 
with the intrinsic ejection velocity of 
$568^{+19}_{-17} \kms$ 
(corrected for the streaming motion of the disk), 
based on the proper motion data from \Gaia\ Data Release 2 (DR2) 
and high-resolution spectroscopy. 
The extremely large ejection velocity indicates that 
this star was not ejected by the supernova explosion of the binary companion. 
Rather, it was probably ejected by a 3- or 4-body dynamical interaction with 
more massive objects in a high-density environment. 
Such a high-density environment 
may be attained at the core region of a young massive cluster with mass of $\gtrsim 10^4~\msun$. 
The ejection agent that took part in the ejection of \hvs\ 
may well be an intermediate mass black hole ($\gtrsim 100~\msun$), 
a very massive star ($\gtrsim 100~\msun$), 
or multiple ordinary massive stars ($\gtrsim 30~\msun$). 
Based on the flight time and the ejection location of \hvs, 
we argue that its ejection agent or its natal star cluster 
is currently located near the Norma spiral arm. 
The natal star cluster of \hvs\ 
may be an undiscovered young massive cluster near the Norma spiral arm. 

\end{abstract}

\section{Introduction}
\label{sec:intro}

Young O- and B-type stars with large peculiar velocities 
($\gtrsim 40 \kms$)  or with large vertical excursion from the Galactic disk plane 
($\gtrsim 1 \kpc$) are called runaway stars.  Some runaway stars are massive 
($\gtrsim 8 M_\odot$; O-type and early B-type), and  are thought to have been ejected 
from their birth places in the stellar disk. The recently published astrometric data from \Gaia\ 
\citep{2018arXiv180409365G} have provided us a unique opportunity 
to accurately reconstruct the orbits of these short-lived massive runaway stars 
and to measure their ejection locations and velocities. Such an analysis enables us not only to investigate  the  processes by which such runaway stars are ejected
but also provide new insights into the environments where massive stars form. 

%\subsection{Two mechanisms for ejecting runaway stars}

The proposed mechanisms for ejecting massive runaway stars 
may be  categorized into two classes \citep{Blaauw1993}: 
\begin{enumerate}
\item binary ejection mechanism (BEM), 
in which a runaway star is ejected as a result of 
the supernova explosion of its binary companion 
\citep{Zwicky1957,Blaauw1961}; and 
\item dynamical ejection mechanism (DEM), 
in which 3- or 4-body interaction of stars (and black holes) 
in high-density environment ejects a runaway star 
\citep{Poveda1967,Aarseth1972,Hut1983}.
\end{enumerate}
In this work we adopt the above nomenclature and acronyms BEM and DEM from \cite{SN2011}.

The most striking difference between these two mechanisms is 
the range of intrinsic ejection velocity 
$V_\mathrm{ej}^\mathrm{int}$
of runaway stars 
with respect to the original binary system (in the case of BEM) 
or the natal star cluster (in the case of DEM).

In the BEM,  the ejection velocity is determined by  the orbital velocity at the time of the supernova explosion 
and the kick velocity arising from the asymmetry of the supernova. 
\cite{Tauris2015} showed that  BEM ejects massive runaway stars of $M=10 M_\odot$
with a velocity up to $V_\mathrm{ej}^\mathrm{int}=320 \kms$.  By taking into account the fact that 
less massive stars can be ejected with larger velocity 
\citep{PortegiesZwart2000},  we estimate that a B-type main-sequence star with $8 M_\odot$---such as \hvs, the subject of this paper---may be ejected with a velocity up to  $V_\mathrm{ej}^\mathrm{int} \sim 400 \kms$ 
($=320 \kms \times 10 M_\odot / 8 M_\odot$), by this mechanism.

For  DEM,  the ejection velocity is determined by  the details of how the binary system is disrupted by the third and/or fourth object,  and the ejection velocity is always smaller than the surface escape velocity from the most massive object involved in the few-body interaction \citep{Leonard1991}.
The surface escape velocity from an object with mass $M_*$ and radius $R_*$ is given by\footnote{
In this paper, both the stellar radius and the Galactocentric radius in the cylindrical coordinate system 
are expressed by $R$. However, the distinction between these quantities is obvious from the context. 
}  
\eq{
v_{\mathrm{esc},*} = 618 \kms 
\sqrt {\frac{M_*}{M_\odot} \frac{R_\odot}{R_*} }, 
}
where $(M_\odot, R_\odot)$ are the mass and radius of the Sun. 
As is obvious from this expression,  this can be enormous for very massive compact objects. 
As a result,  the ejection velocity of a massive runaway star with $M \simeq 8~\msun$
can, in principle, reach 
$V_\mathrm{ej}^\mathrm{int} \sim 10^3 \kms$ 
if a compact object such as a very massive star \citep{Gvaramadze2009} 
participates in the few-body encounter. 
If a black hole participates in the few-body interaction, 
then the above argument is no longer valid 
(the surface escape velocity corresponds to the speed of light). 
However, simulations show that
the typical ejection velocity 
can be as high as $\sim 10^3 \kms$ 
if an intermediate mass black hole (IMBH) 
participates in the few-body interaction 
(\citealt{Gvaramadze2008,Fragione2018}; see also \citealt{Hills1988,YuTremaine2002}).

For both BEM and DEM, 
theoretical studies suggest that 
a large ejection velocity of a runaway star is attained 
when some extreme conditions are met, 
such as very small separation of stars in a binary to be disrupted 
or the existence of a massive compact object 
\citep{Tauris1998,Gvaramadze2008,Gvaramadze2009}. 
Thus, 
massive runaway stars in the Milky Way 
with large ejection velocity 
may provide some insights into the extreme environment 
in which massive stars form 
\citep{Gies1986, Conlon1990, Hoogerwerf2001, deWit2005, Mdzinarishvili2005, Martin2006, SN2011, Tetzlaff2011, Boubert2017, Maiz2018}.

In this regard, the recently discovered \hvs,  a luminous blue star in the inner halo with extremely large velocity \citep{Zheng2014,Huang2017},  is an intriguing object with which to study the extreme conditions 
of the ejection of runaway stars. %under which runaway stars are ejected.
As we will describe briefly in Section \ref{sec:PManalysis}, 
proper motion information for this star from \Gaia\ Data Release 2 (DR2; \citealt{2018arXiv180409365G}) 
as well as its young age suggests that this star was 
not ejected by the supermassive black hole (SMBH) at the Galactic Center (as has been proposed for other hypervelocity stars) but was ejected from the stellar disk.  Due to its (almost) unbound orbital energy, this star can be categorized as a massive hyper-runaway star.  Based on its current velocity, its intrinsic ejection velocity is also expected to be quite large.  This makes \hvs\ a unique star with which we can recover important information about the extreme environments that result in  massive star ejection.

In this paper,  we investigate the origin of this massive hyper-runaway star \hvs\  
by analyzing its orbit and stellar atmosphere 
based on data from \Gaia\ and our spectroscopic observation with Magellan Telescope. 
The outline of this paper is as follows. 
In Section \ref{sec:Gaia_data}, 
we describe the astrometric data from {\Gaia} 
and argue that \hvs\ was ejected from the Galactic disk. %2
In Section \ref{sec:spectroscopic_data},
we derive the stellar parameters from our spectroscopic data 
and show that this star is unambiguously massive. %3
In Section \ref{sec:analysis}, we analyze the orbit of this star.
In Section \ref{sec:result}, we show the results of our analysis. 
In Section \ref{sec:mechanism}, we discuss the possible ejection mechanism of \hvs. 
In Section \ref{sec:rate}, we discuss the ejection rate for each possible ejection mechanism. 
Section \ref{sec:discussion_conclusion} summarizes our conclusions.

\section{Astrometric Data from {\it Gaia-DR2}} \label{sec:Gaia_data}

The astrometric data  for \hvs\ provided by \Gaia\ DR2 \citep{2018arXiv180409365G} 
are summarized in Table \ref{table:stellar_parameters}. 
Here we briefly describe these data and the simple conclusions that can be drawn from them. 
The proper motion of this star is precisely measured but the line-of-sight velocity of this star is not measured by \Gaia. 
The \Gaia-DR2 parallax is 
{\textcolor{black}{
negative and 
}}
associated with a large error, and therefore we do not use this parallax information in our analysis. 
For the distance and the heliocentric line-of-sight velocity,  we use our spectroscopic data as discussed in Section 
\ref{sec:spectroscopic_data}. 
{\textcolor{black}{
The reliability of the astrometric solution 
for this star is discussed in Appendix \ref{sec:Gaia_quality}. 
}}

\subsection{LAMOST-HVS1 is a hyper-runaway star} \label{sec:PManalysis}

Those stars that have been ejected by the SMBH at the Galactic Center 
are often called ``hypervelocity stars'' 
\citep{Hills1988,YuTremaine2002,Brown2015ARAA}. 
Due to the large line-of-sight velocity of \hvs, 
it has been thought 
that this star, which is marginally bound to the Milky Way, may be a hypervelocity star.

\citet{Hattori2018b} showed that for a star at the distance of \hvs\ 
($d \sim 10$-$20 \kpc$), 
the azimuthal angular momentum $L_z$ has to be extremely close to zero if it is ejected by the SMBH (since the potential of the Milky Way in this region is close to axisymmetric). They showed that if \hvs\ was ejected from the Galactic Center its proper motion would have to lie in a very narrow range of values (see  Figure 4(a) of \citealt{Hattori2018b}). However, \Gaia's high-precision proper motion measurements 
$(\mualpha, \mudelta) = (-3.54 \pm 0.11, -0.62 \pm 0.09) \masyr$ of \hvs\ indicate that its observed proper motions are so different from those predicted  in the Galactic-Center-origin scenario that these data alone rule out this mechanism.

Furthermore, the young age of this star, already inferred from LAMOST spectra \citep{Zheng2014,Hattori2018b},  combined with the above-mentioned simple analysis of the proper motion, suggests that \hvs\ was born in the Galactic disk and was ejected to reach its current location several $\kpc$ away from the disk plane.  
Given its large current velocity, this star is almost unbound  to the Milky Way and thus \hvs\ is a hyper-runaway star.

\section{Spectroscopic data} \label{sec:spectroscopic_data}

%\startlongtable
\begin{deluxetable}{c c}
\tablecaption{Basic properties of LAMOST-HVS1 
\label{table:stellar_parameters}}
%\rotate %%Kohei
\tablewidth{0pt}
\tabletypesize{\scriptsize}
\tablehead{
\colhead{{\it Gaia} data } &
\colhead{LAMOST-HVS1} 
}
\startdata
    {\it Gaia} DR2 source\_id & $590511484409775360$ \\
	$\ell$ & $221.099505130^\circ$ \\
	$b$ & $ 35.407214626^\circ$ \\
	$\alpha$ & $138.027167145^\circ$ \\
	$\delta$ & $9.272744019^\circ$ \\
	%$B$ & $12.936 \pm 0.036$ \\
	%$V$ & $13.055 \pm 0.009$ \\
	%$J$ & $13.357 \pm 0.028$ \\
	%$H$ & $13.43 \pm 0.04$ \\
	%$K_s$ & $13.53 \pm 0.04$ \\
	$G/\mathrm{mag}$ & $13.06$ \\ %$13.063512$
	\textcolor{black}{$(G_\mathrm{BP} - G_\mathrm{RP})/\mathrm{mag}$} & \textcolor{black}{$-0.2316$} \\
    $\varpi / \mas$ & $-0.044$ \\ % $-0.04413743560939153$
    $\sigma_\varpi/ \mas$ & $0.067$ \\ %$0.06688629816963644$  
    $\mualpha / (\masyr)$ & $-3.54$ \\ %$-3.5374250276479575$
    $\sigma_{\mu,\alpha*}/ (\masyr)$ & $0.11$ \\ %$0.10988840715270515$
    $\mudelta/ (\masyr)$ & $-0.62$ \\ %$-0.6196247281732981$
    $\sigma_{\mu,\delta}/ (\masyr)$ & $0.09$ \\%$0.09301385825953493$ \\
	\hline\hline
	Spectroscopic data & $\;$ \\
	\hline
	$\teff$ &  $18100 \pm 400 \;\mathrm{K}$ \\ 
	$\log (g)$ & $3.42 \pm 0.065$ \\ 
	$\vlos$ & $615 \pm 5 \kms$ \\
	$v \sin (i)$ &  \textcolor{black}{$130 \pm 20 \kms$} \\ 
	$v_\mathrm{macro}$ &  \textcolor{black}{$164 \pm 30 \kms$} \\ 
	$v_\mathrm{micro}$ &  \textcolor{black}{$6 \pm 1 \kms$} \\ 
	age $\tau$ & $37.4^{+4.0}_{-3.7} \Myr$\\ %$37.373^{+3.957}_{-3.723} \Myr$  \\
	(Current mass) $M_{*}$ & $8.3^{+0.5}_{-0.4} M_\odot$\\ %$8.308^{+0.465}_{-0.369} M_\odot$ \\
	(Current radius) $R_{*}$ & $9.3^{+1.0}_{-0.9} R_\odot$\\ %$9.314^{+1.032}_{-0.923} R_\odot$  \\	
	$DM$ & $16.40^{+0.52}_{-0.48}$ \\%$16.404 ^{+0.518}_{-0.481}$ \\
	$d_\mathrm{spec}$ & $19.1^{+5.1}_{-3.8} \kpc$ \\ %$19.094^{+5.140}_{-3.793} \kpc$ \\
	\hline
	\hline
	Result of orbital analysis & $\;$ \\
    \hline
    $d$ & $13.3^{+1.7}_{-1.5} \kpc$ \\% $d$ & $13.38^{+1.70}_{-1.48} \kpc$ \\
\enddata
\tablecomments{
The posterior distribution of the distance 
after the orbital analysis is denoted as $d$, 
while the spectroscopic distance used as an input of the orbital analysis 
is denoted as $d_\mathrm{spec}$ 
(see Figure \ref{fig:hist_distKin_distSpec}).
}
%\tablenotetext{a}{}
\end{deluxetable}

%\startlongtable
\begin{deluxetable}{l c l l c c c c c}
\tablecaption{Chemical abundances of LAMOST-HVS1 
\label{table:abundances}}
%\rotate %%Kohei
\tablewidth{0pt}
\tabletypesize{\scriptsize}
\tablehead{
\colhead{} &
\colhead{} &
\multicolumn{2}{c}{LAMOST-HVS1} &
\colhead{} &
\multicolumn{2}{c}{B stars} &
\colhead{} &
\colhead{Sun} \\
\cline{3-4}
\cline{6-7}
\cline{9-9}
\colhead{X} &
\colhead{} &
\colhead{[X/H]$^{\mathrm{(a)}}$} &
\colhead{A(X)} &
\colhead{} &
\colhead{$\langle\mathrm{A}_\mathrm{B}^{3 \kpc} (\mathrm{X})\rangle$} &
\colhead{$\langle\mathrm{A}_\mathrm{B}^{8 \kpc} (\mathrm{X})\rangle$} &
%\colhead{A$_\mathrm{B}^{3 \kpc}$(X)} &
%\colhead{A$_\mathrm{B}^{8 \kpc}$(X)} &
\colhead{} &
\colhead{A$_\odot$(X)} 
}
\startdata
Si & & $0.60\pm0.06$ & $8.11^{+0.05}_{-0.05}$	& &$7.73\pm0.05$ &$7.50\pm0.05$ & &$7.51\pm0.03$ \\ 
Mg & & $0.33\pm0.10$ & $7.93^{+0.06}_{-0.12}$	& &$7.76\pm0.05$ &$7.56\pm0.05$ & &$7.60\pm0.04$ \\ 
C  & & $0.26\pm0.07$ & $8.69^{+0.05}_{-0.06}$	& &$8.85\pm0.10$ &$8.33\pm0.04$ & &$8.43\pm0.05$ \\ 
N  & & $0.35\pm0.07$ & $8.18^{+0.05}_{-0.05}$	& &$8.22\pm0.11$ &$7.79\pm0.04$ & &$7.83\pm0.05$ \\ 
O  & & $0.17\pm0.07$ & $8.86^{+0.05}_{-0.05}$	& &$8.94\pm0.05$ &$8.76\pm0.05$ & &$8.69\pm0.05$ \\ 
\enddata
\tablecomments{
A(X) denotes the abundance of \hvs\ derived from our analysis.
{\textcolor{black}{
$\langle\mathrm{A}_\mathrm{B}^{3 \kpc} (\mathrm{X})\rangle$ and 
$\langle\mathrm{A}_\mathrm{B}^{8 \kpc} (\mathrm{X})\rangle$ 
denote the mean abundance of the B stars 
located at $R=3 \kpc$ and $8\kpc$, respectively, 
derived from table 9 of \cite{Nieva2012} 
and the literature values of the radial abundance gradient therein. 
%\citep{Carigi2005,Esteban2005,Cescutti2007}. 
}}
A$_\odot$(X) denotes the abundance of the Sun taken from \cite{Asplund2009ARAA}.
{\textcolor{black}{
$^{\mathrm{(a)}}$Note that 
[X/H]$=\mathrm{A(X)- A_\odot(X)}$  
depends not only on $\mathrm{A(X)}$ but also on an external quantity $\mathrm{A_\odot(X)}$. 
Thus, any systematic error on $\mathrm{A_\odot(X)}$ can affect [X/H]. 
}} 
}
%\tablenotetext{a}{}
%\startdata
%Si & $0.60$ & $8.11^{+0.05}_{-0.05}$	&$7.51\pm0.03$ \\ 
%Mg & $0.33$ & $7.93^{+0.06}_{-0.12}$	&$7.60\pm0.04$ \\ 
%C  & $0.26$ & $8.69^{+0.05}_{-0.06}$	&$8.43\pm0.05$ \\ 
%N  & $0.35$ & $8.18^{+0.05}_{-0.05}$	&$7.83\pm0.05$ \\ 
%O  & $0.17$ & $8.86^{+0.05}_{-0.05}$	&$8.69\pm0.05$ \\ 
%\enddata
\end{deluxetable}

\subsection{Observation of LAMOST-HVS1} 

We observed \hvs\ on 2018 January 12 using the 
Magellan Inamori Kyocera Echelle (MIKE)
spectrograph \citep{bernstein03} mounted on the
6.5~m Landon Clay Telescope (Magellan~II)
at Las Campanas Observatory, Chile.
The 1\farcs0$\times$5\farcs0 entrance slit
and 2$\times$2 detector binning yielded
spectral resolving powers of 
$R \equiv \lambda/\Delta\lambda \sim$ 
30,000 and 25,500
on the blue and red spectrographs, respectively.
Light entering the two spectrographs is split by a dichroic
at $\approx$~4950~\AA.~
Observations were split into 7 sub-exposures,
and the total integration time was 1.7~hr.
The spectra cover 
3350~\AA$\leq \lambda \leq$~9400~\AA,
although the spectra redward of $\approx$~8300~\AA\
suffer from fringing.
The spectra were reduced using the CarPy MIKE pipeline
\citep{kelson00,kelson03}, including 
overscan subtraction, flat field division, image co-addition,
cosmic ray removal, sky and scattered light subtraction, 
rectification of the tilted slit profiles,
spectrum extraction, and wavelength calibration
based on ThAr comparison lamp spectra
taken between science exposures.
Final signal-to-noise (S/N) ratios per pixel range from
$\approx$~50 at 3500~\AA,
$\approx$~200 at 4500~\AA,
$\approx$~140 at 5500~\AA, to
$\approx$~150 at 6500~\AA.~
These spectra can be distributed upon request.

%%%%% Figure 1 %%%%%
\begin{figure*}
\begin{center}
\includegraphics[angle=0,width=7.in]{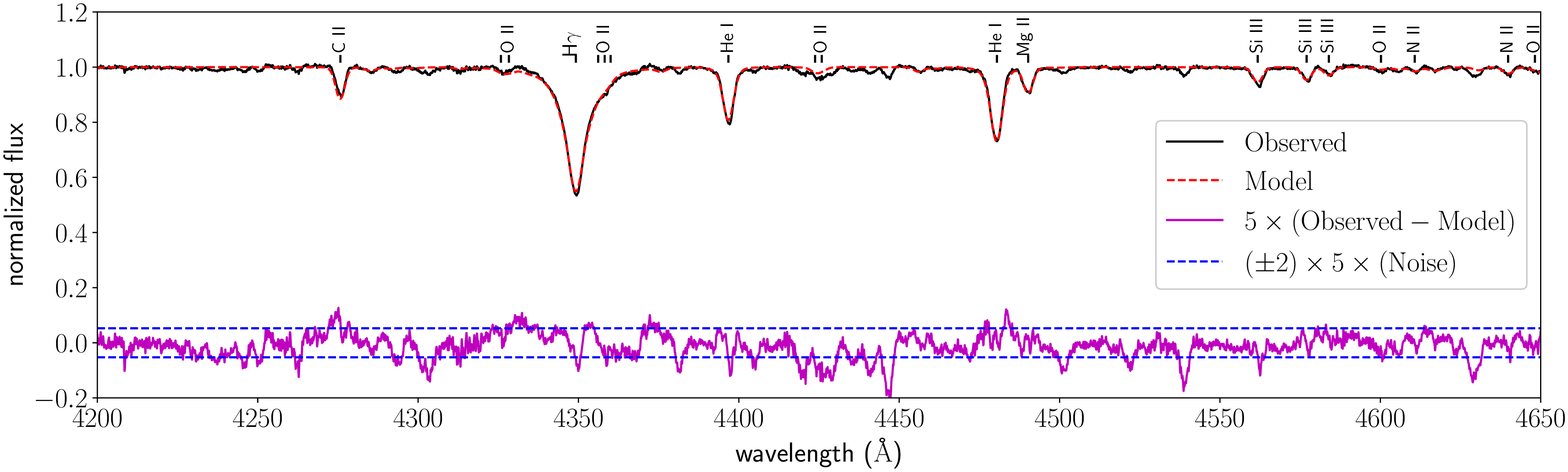} 
\includegraphics[angle=0,width=7.in]{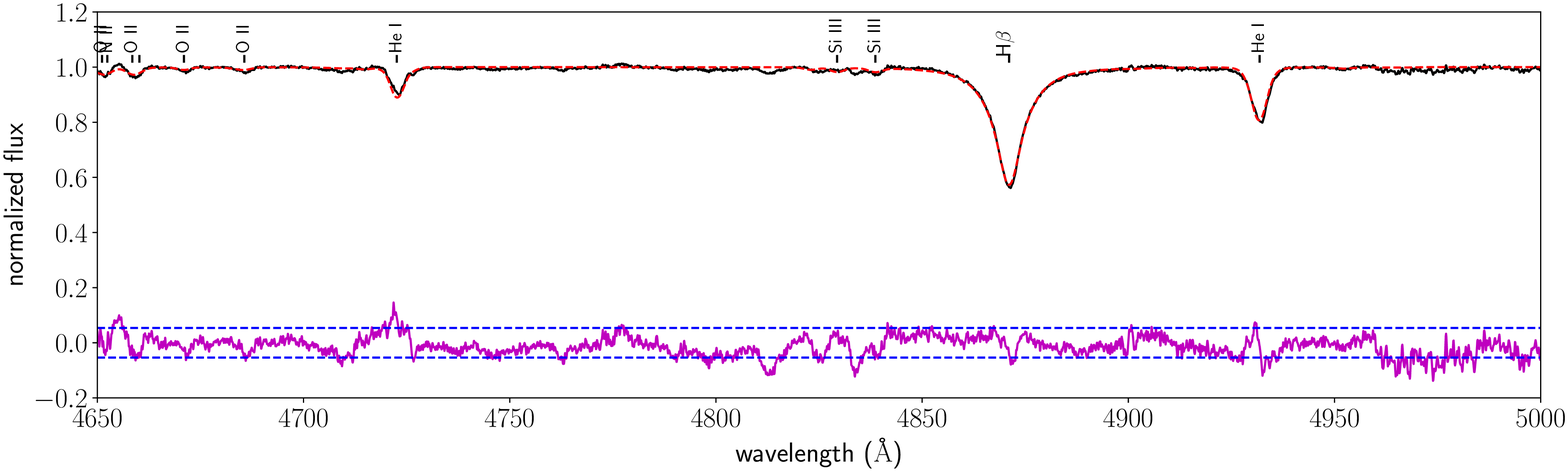} 
\includegraphics[angle=0,width=7.in]{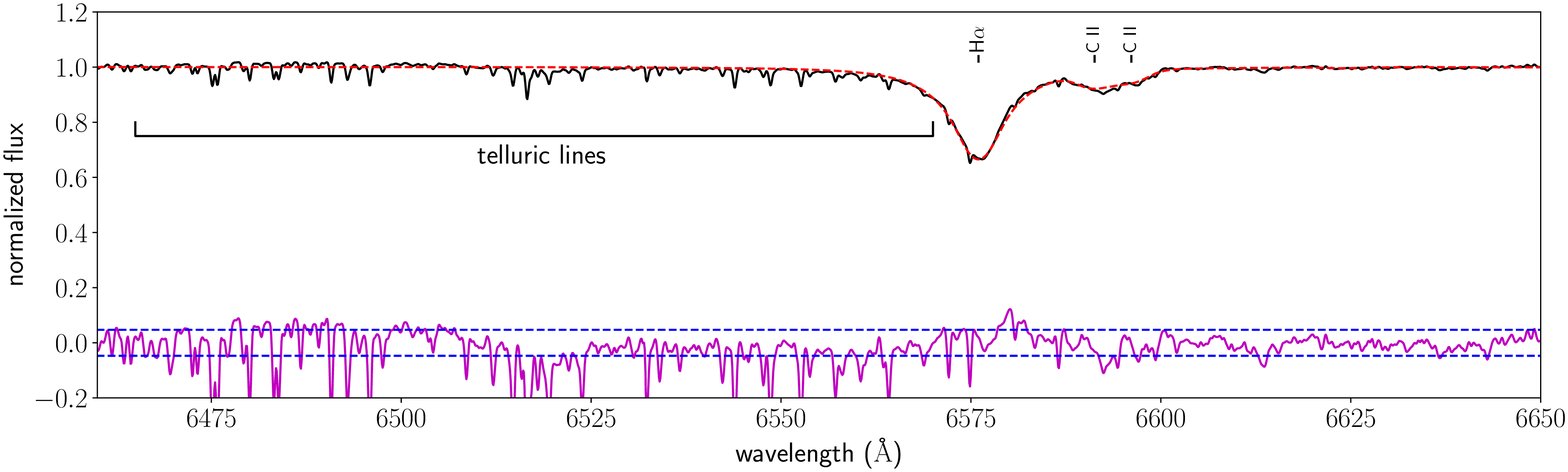} 
\end{center}
\caption{
\label{fig:mike}
The normalized MIKE spectrum of \hvs\ (black solid line) 
with some absorption lines annotated. 
The red dashed line is the synthetic {\sc fastwind} model 
using the best-fit parameters in 
Tables \ref{table:stellar_parameters} and \ref{table:abundances}.  
We note that the helium abundance is fixed to approximate Solar abundance. 
{\textcolor{black}{
The residual spectrum (enlarged by a factor of 5) is 
shown by the magenta solid line, 
and the $2\sigma$ noise level (enlarged by a factor of 5) 
is shown by the blue dashed lines. 
}}
The high signal-to-noise ratio of our spectroscopic data and 
the good match with the model  
demonstrate the reliability of our stellar parameters and abundances. 
{\textcolor{black}{
We note that the observed spectrum has not been shifted to the rest-frame. 
(That is, we do not correct 
the observed spectrum 
for 
the heliocentric line-of-sight velocity, $\vlos = 615 \kms$, 
and the observer's motion relative to the Sun, $-13 \kms$; 
we shift our model spectrum instead.) 
Analyzed metallic absorption lines are shown (by vertical line segment) 
only when the equivalent width 
is larger than $0.025$~\AA. 
Some absorption lines that are shown (by vertical line segment) 
but not annotated 
(e.g., those on both sides of H$\gamma$ line) 
are $\mathrm{O_{II}}$ lines. 
The sharp absorption features left to H$\alpha$ are telluric lines.
}}
}
\end{figure*}

\subsection{Analysis of spectra of LAMOST-HVS1} 

The stellar atmosphere analysis is based 
on a grid of synthetic atmosphere models built using the atmosphere/line formation code {\sc fastwind} \citep{1997A&A...323..488S,2005A&A...435..669P,2012A&A...543A..95R}. The projected rotational  (\ensuremath{{\upsilon}\sin i}) and macroturbulence velocities are estimated using the {\sc iacob-broad} code \citep{2007A&A...468.1063S,2014A&A...562A.135S}. 

The modeling of the data is carried out in two steps. 
Initially, the observed \hvs\ spectrum is compared with the synthetic {\sc fastwind} grid, retrieving the set of stellar parameters that best reproduce the main 
%chemical transitions 
absorption lines %[KH]
in the data. 
{\textcolor{black}{
The errors in the macroturbulence and rotational velocities are estimated using the Fourier approach and goodness-of-fit algorithms described in 
\cite{2007A&A...468.1063S,2014A&A...562A.135S}. 
}}
%2017A&A...597A..22S
Based on the best model, a new grid is designed exploring different chemical compositions using fine steps of 0.1\,dex in the abundances  of the  elements displayed in the observations (i.e. Si, Mg, C, N and O)  
{\textcolor{black}{
and, microturbulence in steps of $1 \kms$. 
}}
Also, two abundance values for helium, namely $Y=$ He/H $=0.10$ (approximate Solar abundance) and $0.15$ (super-Solar abundance), are tried; and we fix it to the Solar abundance as it results in a better fit to the data. 
The observed spectrum is cross-matched with this new grid until we find the composition 
{\textcolor{black}{
and microturbulence that are 
}}
able to reproduce the different spectral features. 
A detailed description of the technique and main transitions used in the analysis can be found in \cite{2012A&A...542A..79C} \citep[see also][]{2010A&A...515A..74L}. 
The stellar parameters and chemical composition 
{\textcolor{black}{
(and their $1\sigma$ random error)
}}
obtained from the quantitative analysis 
are listed in 
Tables \ref{table:stellar_parameters} and \ref{table:abundances}. 
The best-fit synthetic spectrum and the observed spectrum  
are shown in Figure \ref{fig:mike}.

{\textcolor{black}{
It is worth mentioning that  
we obtain a large macroturbulence from our analysis. 
\cite{2017A&A...597A..22S} also found stars with large macroturbulence at similar mass ranges. 
Our estimation is indeed higher than their findings in the {\sc iacob} sample. 
Below $15 \msun$ the authors suggest different mechanisms that could induce this additional broadening: the presence of stellar spots, surface inhomogeneities, 
and stellar oscillations \citep{2014A&A...569A.118A}. 
}}

The heliocentric line-of-sight velocity of \hvs\ 
derived from our spectroscopy is $\vlos = 615 \pm 5 \kms$, 
which is consistent with the previous measurements by LAMOST, 
$611.65\pm4.63 \kms$ \citep{Huang2017}. 
This agreement 
{\textcolor{black}{
probably suggests 
}}
that this star is not in a binary system.

The stellar distance is estimated in the following manner. 
First, we draw 1000 samples of $(\teff, \logg)$ from the error distribution mentioned above. 
Then, for each realization of $(\teff, \logg)$, we make a point-estimate of the stellar radius $R_{*}$ based on the rotating evolutionary tracks published by \cite{2012A&A...537A.146E}. 
For each set of $(\teff, \logg, R_{*})$ and the corresponding synthetic {\sc fastwind} spectral energy distribution, 
we calculate the probability distribution function (PDF) of the distance given the observed optical and IR photometry of \hvs\ 
{\textcolor{black}{
($B$ and $V$ 
from APASS DR9 \citealt{Henden2016}; 
$J$, $H$, $K_s$ 
from 2MASS \citealt{Cutri2003}; 
all of these quantities are listed in table 1 of \citealt{Hattori2018b}).
}}
We adopt the standard \cite{1989ApJ...345..245C} extinction law 
{\textcolor{black}{
with $R_V=3.1$; 
and the adopted color excess is 
$E(B-V)=0.085$ for this star. 
}}
Finally, we add these 1000 PDFs to obtain the full PDF of the spectroscopic distance $d_\mathrm{spec}$. 
\Gaia\ reports a negative parallax $\varpi$ for \hvs, 
but the median value of our spectroscopic distance is consistent with 
\Gaia's measurement within two standard deviations level 
($|\varpi - 1/d_\mathrm{spec,median}| < 2 \sigma_\varpi$).

\subsection{Is LAMOST-HVS1 a massive star?}

In general, heavier runaway stars 
are more difficult to accelerate to  large velocities 
since they need to receive larger momentum \citep{PortegiesZwart2000}. 
Thus, when studying the ejection mechanisms for  a runaway star,  it is important to accurately determine its stellar type (and hence its mass)
\citep{SN2011,McEvoy2017}.

Our analysis suggests that 
\hvs\ is a massive subgiant star with $M_* \simeq 8 M_\odot$. 
In order to validate our stellar classification, 
we provide additional lines of  evidence (Section~\ref{sec:notBHB})
and argue that \hvs\ is not a hot blue horizontal branch star 
(BHB star; $\lesssim 1 M_\odot$). 
We note that some horizontal branch stars 
extended beyond $\teff > 20000 \Kelvin$ 
are often classified as hot subdwarfs 
\citep{Greenstein1974, Heber2009ARAA}; 
however the detailed definition of BHB stars 
is not important in this paper.

\subsubsection{Evidence that LAMOST-HVS1 is not a BHB star}
\label{sec:notBHB}

Our analysis suggests that \hvs\ shows 
$(\teff, \logg) = (18100 \pm 400 \Kelvin, 3.42 \pm 0.065)$. 
These values are inconsistent with a BHB star. 
The surface gravity for a BHB star as hot as \hvs\ 
is about $\logg \simeq 4.6 \pm 0.2$, 
while a BHB star with $\logg \simeq 3.42$ 
has $\teff \simeq 10000 \Kelvin$ 
\citep{Dorman1993,MoniBidin2007,MoniBidin2011}.

Our measurement of large projected rotational velocity, $v \sin (i) = 130 \kms$, 
of \hvs\ is typical of massive stars. 
In contrast, hot BHB stars with $\teff > 11500 \Kelvin$ 
show slow rotation with
$v \sin (i) < 7 \kms$ 
\citep{Behr2000}.

Also, a hot BHB star with $\teff > 11500 \Kelvin$ 
shows a depleted helium abundance of $Y=$ He/H $\lesssim 10^{-2}$ 
\citep{Behr1999, MoniBidin2007}, 
probably due to the diffusion (gravitational settling) of helium 
in the atmosphere \citep{Greenstein1967,Michaud1983,Michaud2015}. 
In contrast, the helium abundance of \hvs\ is consistent with Solar abundance.

Furthermore, the effective temperature and the metal abundance 
of this star are inconsistent with the assumption that this star is a BHB star. 
If this is a BHB star, 
then the intrinsic overall metallicity [M/H] 
has to be as metal-poor as [M/H]$<-1$ \citep{Xue2008}, 
since a metal-rich horizontal branch star 
is necessarily red \citep{Sandage1960,Lee1994,Gratton2010}. 
In contrast, the intrinsic overall metallicity 
estimated from [Mg/H] and [Si/H], 
which are insensitive to $\teff$ or 
the internal evolution of a horizontal branch star 
\citep{Glaspey1989,Behr1999,Behr2003,Moehler2003,Gratton2004ARAA,Fabbian2005},
is super-Solar 
([Mg/H]$=0.33$ and [Si/H]$=0.60$ if we adopt 
the Solar abundance from \citealt{Asplund2009ARAA}). 
Therefore it is highly unlikely that this star is a BHB star.

Based on these considerations, we claim that \hvs\ 
is indeed a massive B-type subgiant star, 
confirming previous claims 
\citep{Zheng2014, Huang2017, Hattori2018b}.

%%%%% Figure 2 %%%%%
\begin{figure*}
\begin{center}
\includegraphics[angle=0,width=5.3in]{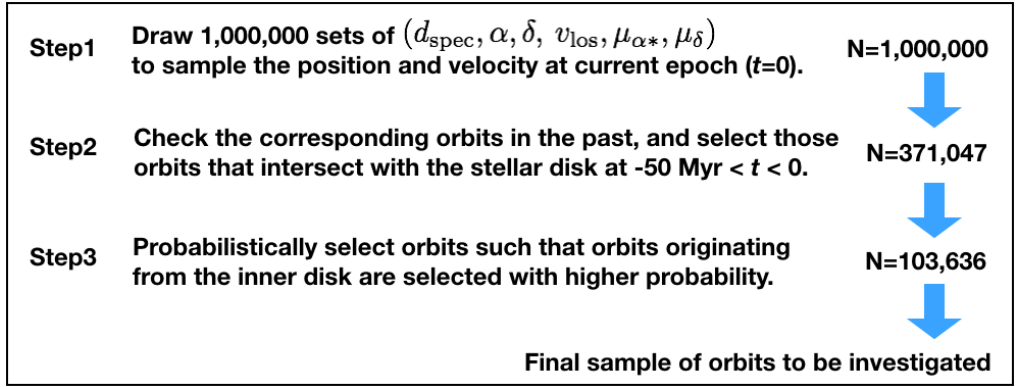} 
\end{center}
\caption{
\label{fig:flow_chart}
A flow chart that describes 
how we construct sample orbits from the error distribution.}
\end{figure*}

%\section{Reconstruction of the orbit of LAMOST-HVS1}
\section{Analysis of the orbit of LAMOST-HVS1}
\label{sec:analysis}

\subsection{Coordinate system and the model potential}
\label{sec:model_potential}

We adopt a Galactocentric inertial right-handed Cartesian coordinate system $(x,y,z)$,
such that $(x,y)$-plane is the Galactic disk plane and 
$z$-axis is directed toward the North Galactic Pole. 
We assume that the Sun is on the Galactic plane $z=0$ 
\citep{Bovy2017}, 
and that the position of the Sun is $(x,y)=(-R_0, 0)$
with $R_0 = 8.0 \kpc$. 
Also, we define a Galactocentric cylindrical coordinate system 
$(R, \phi, z)$ such that $(x,y) = (R \cos \phi, R \sin \phi)$. 

We assume that the circular velocity at the Solar position is
$v_0 = 220 \kms$ \citep{Kerr1986} 
and the Solar peculiar velocity relative to the 
circular velocity is 
$(U_\odot, V_\odot, W_\odot) =$ $(11.1, 12.24, 7.25) \kms$ 
\citep{Schonrich2010}. 
We adopt a realistic Galactic potential model, 
\texttt{MWPotential2014} \citep{Bovy2015}. 
We note that \texttt{MWPotential2014} model 
uses a spherical NFW dark matter halo 
with scale length $a=16 \kpc$,
virial radius $r_{200}=183 \kpc$, and 
virial mass $M_{200}=0.7\times 10^{12} ~\msun$. 
However, the result of this paper is not sensitive to 
the adopted virial mass $M_{200}$, 
as long as $a$ and $r_\mathrm{vir}$ are tuned so that the rotation curve is almost unchanged. 
For example, by taking into account that 
recent measurements of $M_{200}$ of the Milky Way 
based on \Gaia\ DR2 
(e.g., \citealt{posti_helmi2018,Watkins2018};
see also \citealt{Hattori2018c})
is $\sim 2$ times more massive than that of \texttt{MWPotential2014}, 
we have checked that the result of this paper is unchanged 
when we adopt 
$(a, r_{200}, M_{200})=(28 \kpc, 231 \kpc, 1.4\times10^{12} ~\msun)$.

%%%%% Figure 3 %%%%%
\begin{figure*}
\begin{center}
\includegraphics[angle=0,width=5.in]{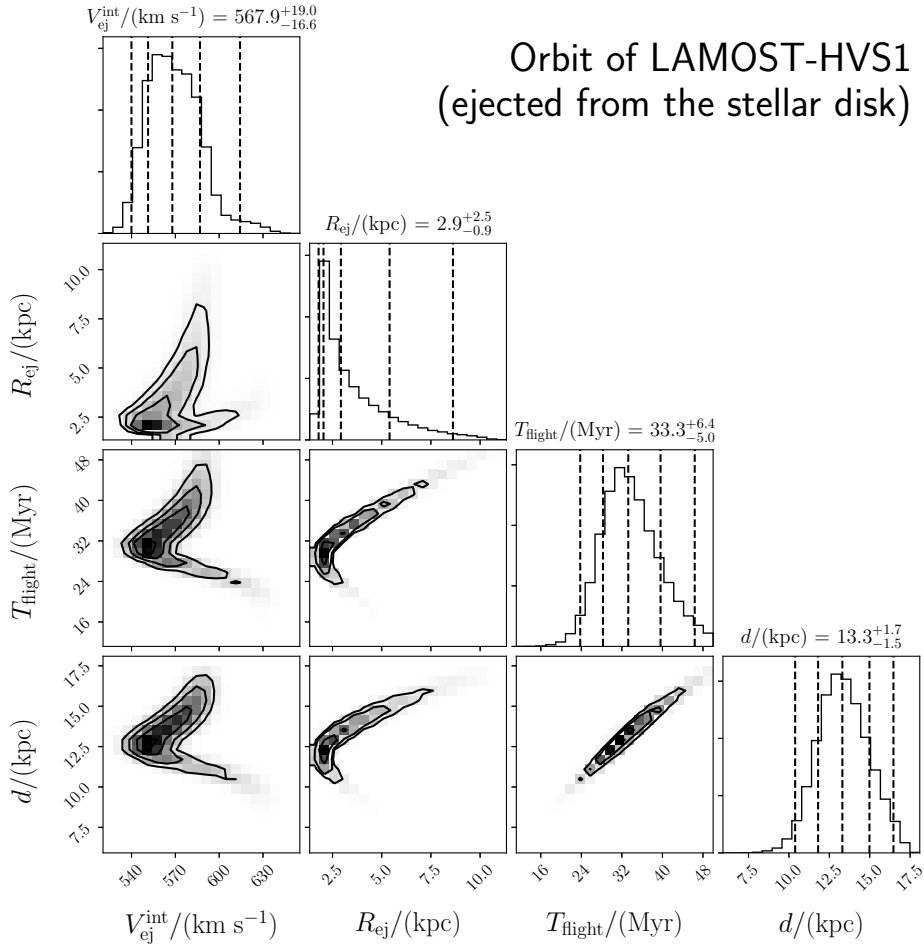} 
\end{center}
\caption{
\label{fig:triangle_Vej_Rej_flightTime_dist}
The probability distribution of the orbital properties of \hvs\ 
represented by 103,636 orbits sampled from our analysis. 
The uncertainty in these orbital parameters are 
almost exclusively explained by the distance uncertainty, 
as described in Section \ref{sec:distance}. 
The dashed vertical lines in the histograms 
represent the 2.5th, 16th, 50th, 84th, and 97.5th percentiles of the distributions. 
}
\end{figure*}

%%%%% Figure 4 %%%%%
\begin{figure}
\begin{center}
\includegraphics[angle=0,width=3.in]{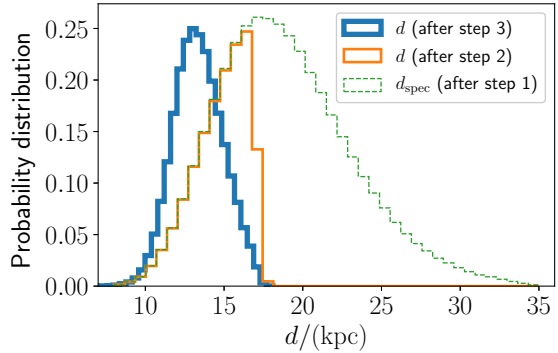} 
\end{center}
\caption{
\label{fig:hist_distKin_distSpec}
The probability distribution of the distance to \hvs. 
The distance distribution of the sampled orbits 
after (step 1), (step 2), and (step 3) in Figure \ref{fig:flow_chart} 
are shown by the green dashed, orange thin, and blue thick histograms,
respectively. 
We see that the spectroscopic distance (green dashed histogram) 
is comparatively uncertain by itself, 
and our belief in the distance 
is improved after we take into account the 
orbital information (blue solid histogram). 
The blue thick and orange thin histograms are normalized to unity; 
while the green dashed histograms is scaled so that the difference 
from the orange thin histogram can be clearly seen.
}
\end{figure}

%%%%% Figure 5 %%%%%
\begin{figure}
\begin{center}
\includegraphics[angle=0,width=1.6in]{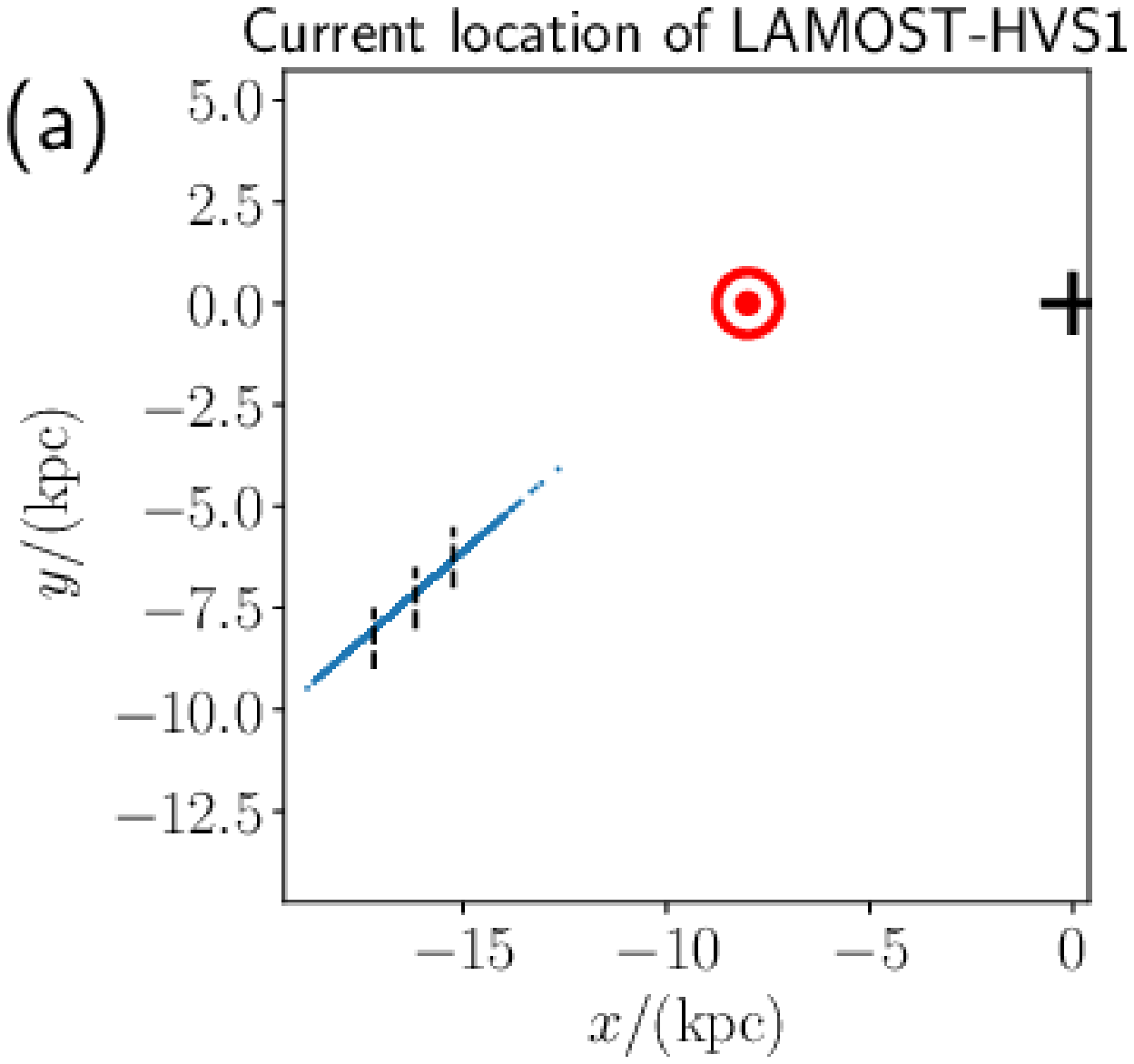} 
\includegraphics[angle=0,width=1.6in]{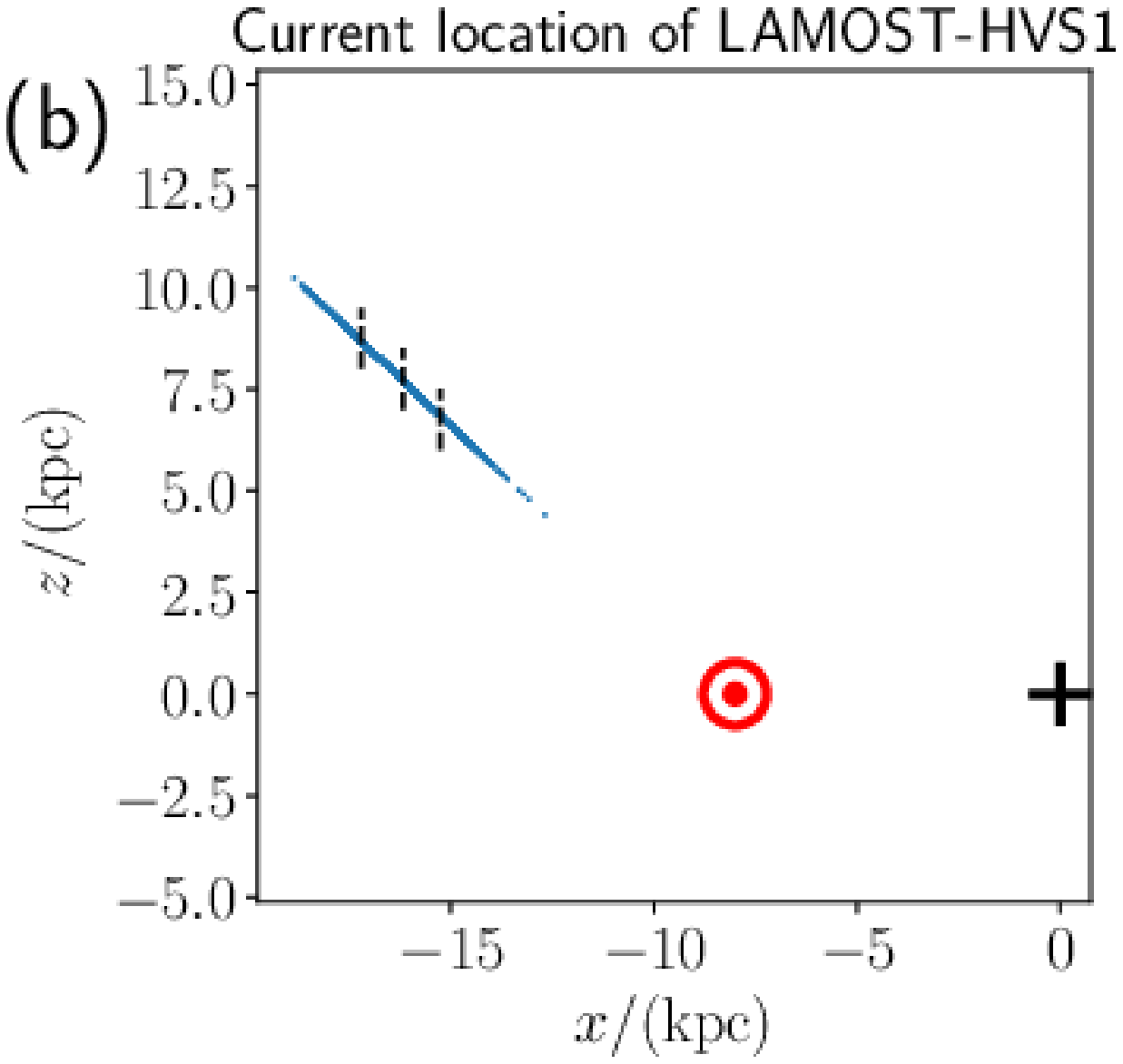} \\
\includegraphics[angle=0,width=3.in]{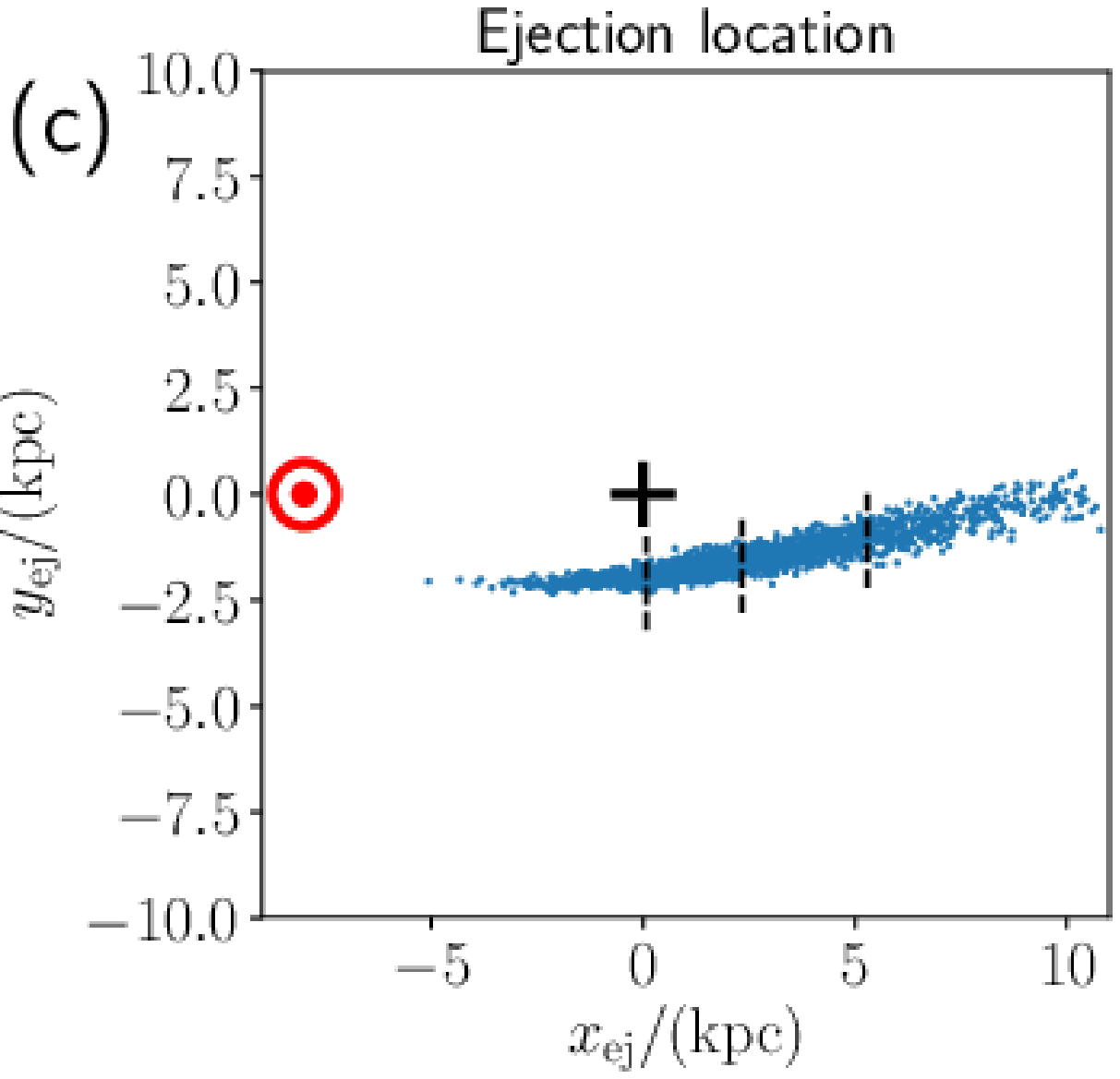}
\end{center}
\caption{
\label{fig:xy}
The probability distribution of the current location of \hvs\
(upper two panels)
and the location when it was ejected from the stellar disk
(bottom panel)  
represented by the 103,636 sample orbits. 
{\textcolor{black}{
(Only $2\%$ of them are shown for clarity.) 
}}
The black plus and red dot correspond to the Galactic Center 
and the current Solar position, respectively. 
In panels (a) and (b), 
the vertical dashed lines indicates the 16th, 50th, and 84th percentiles of 
the current $x$ coordinate of \hvs, 
which are shown to provide a rough idea of the distance uncertainty. 
In panel (c), 
the vertical dashed lines indicate the 16th, 50th, and 84th percentiles of 
the $x$ coordinate of the ejection location, $x_\mathrm{ej}$. 
}
\end{figure}

%%%%% Figure 6 %%%%%
\begin{figure}
\begin{center}
\includegraphics[angle=0,width=3.3in]{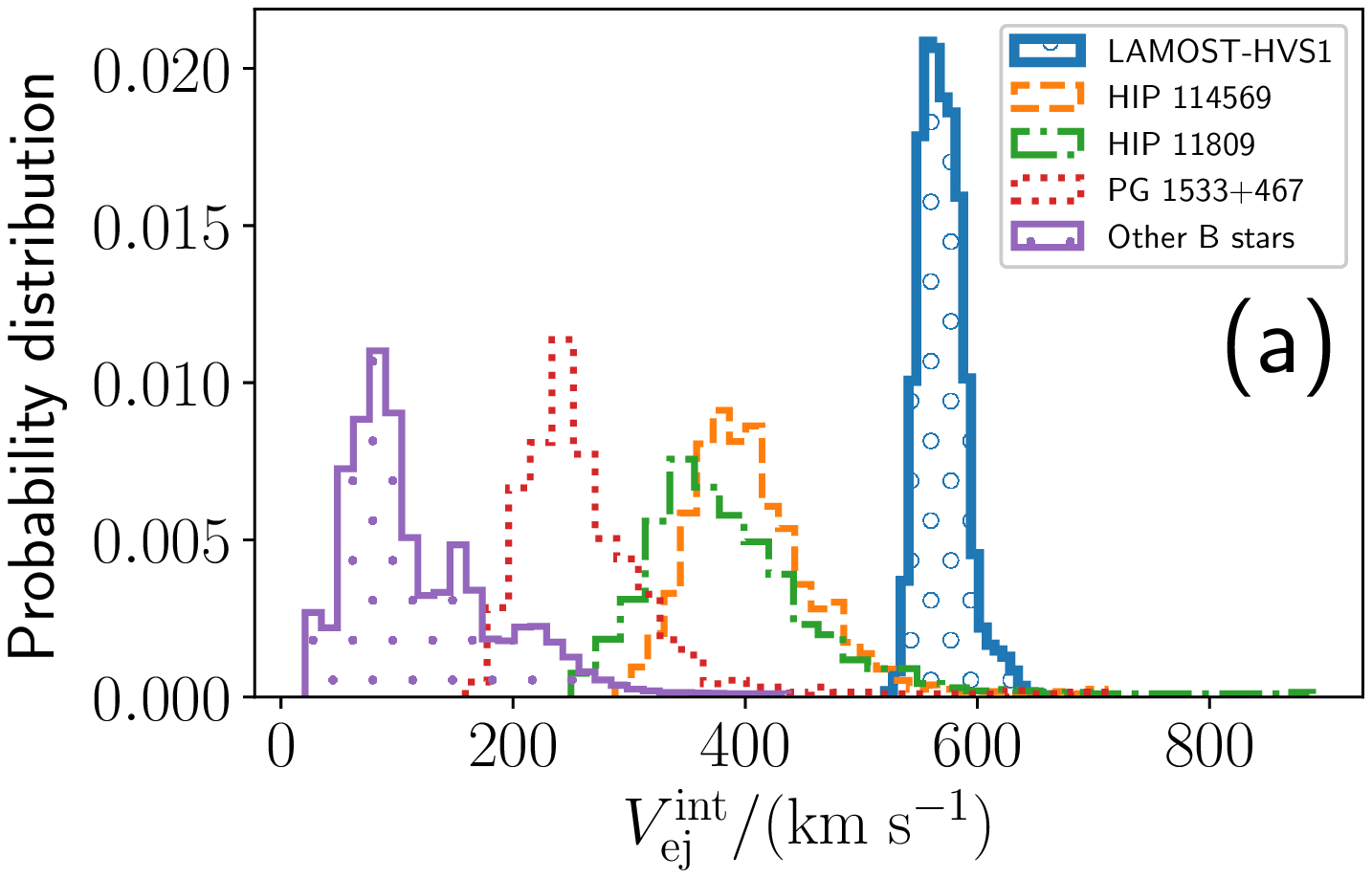} \\
\includegraphics[angle=0,width=3.3in]{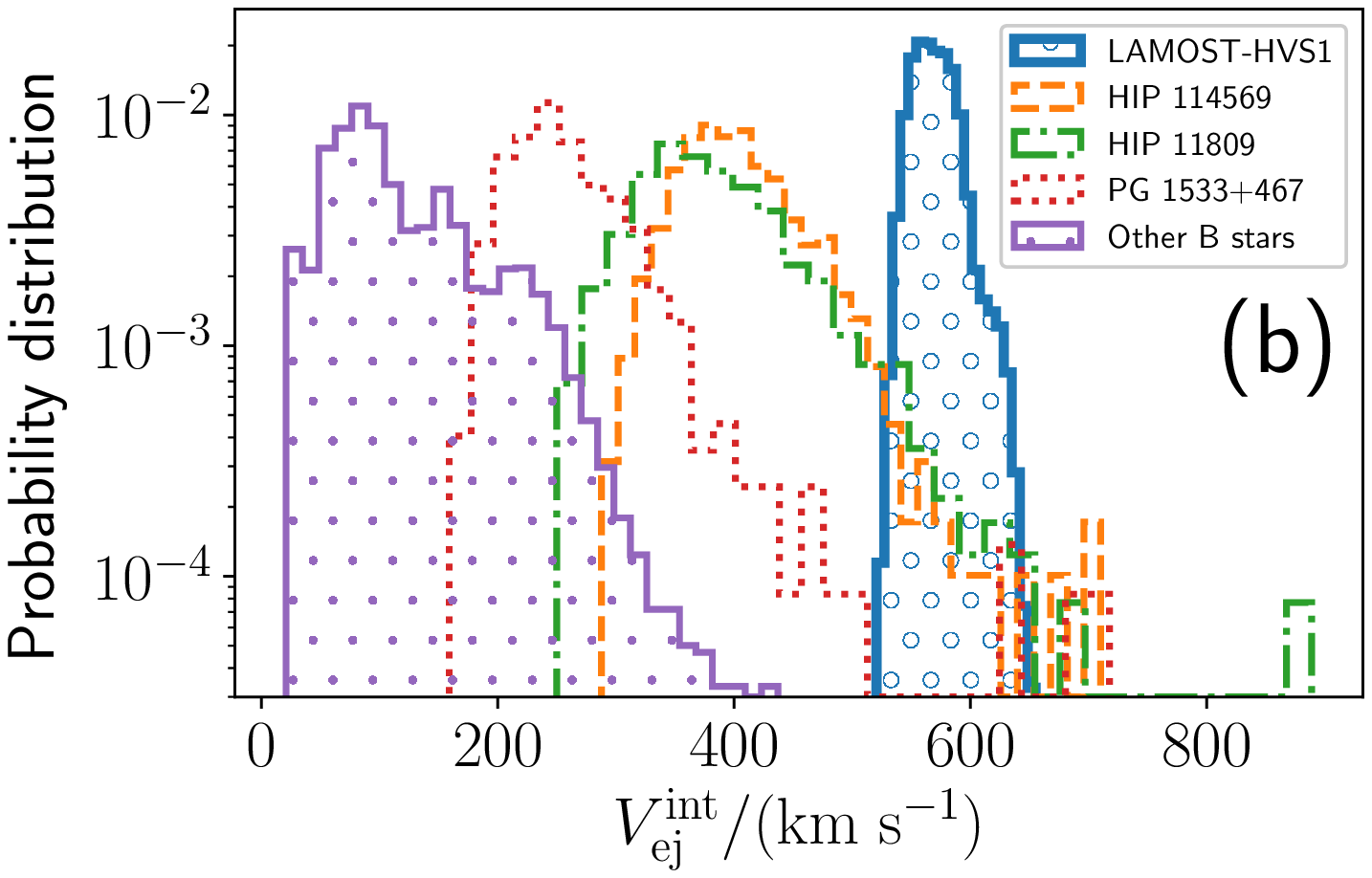} 
\end{center}
\caption{
\label{fig:hist_vej}
The probability distribution of the ejected velocity 
for \hvs\ and 46 other runaway stars in \cite{SN2011} with good \Gaia\ parallaxes.
For clarity, we show the same histograms
in linear and logarithmic scales
on top and bottom panels, respectively. 
}
\end{figure}

%%%%% Figure 7 %%%%%
\begin{figure}
\begin{center}
\includegraphics[angle=0,width=3.in]{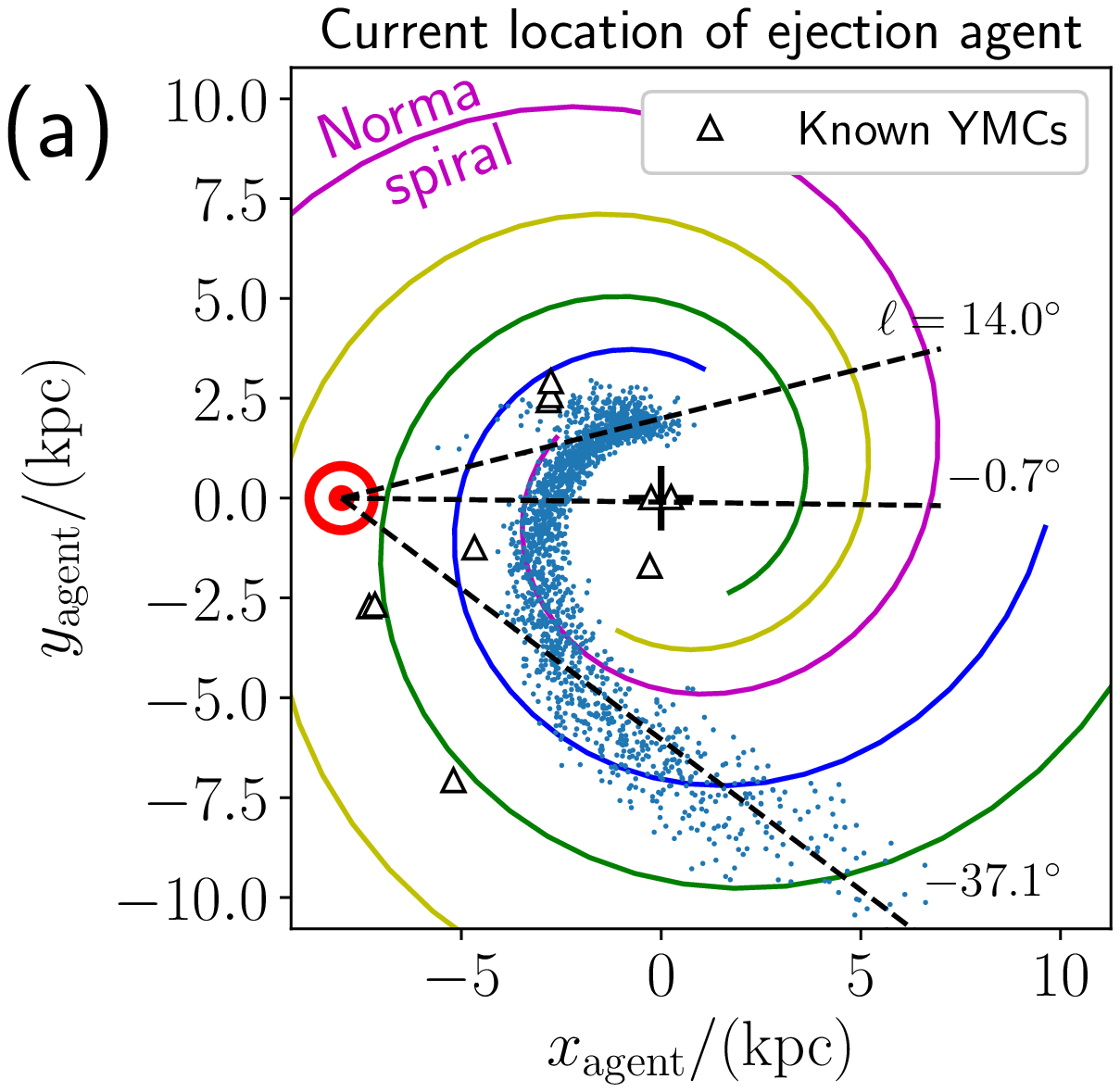} \\
\includegraphics[angle=0,width=3.in]{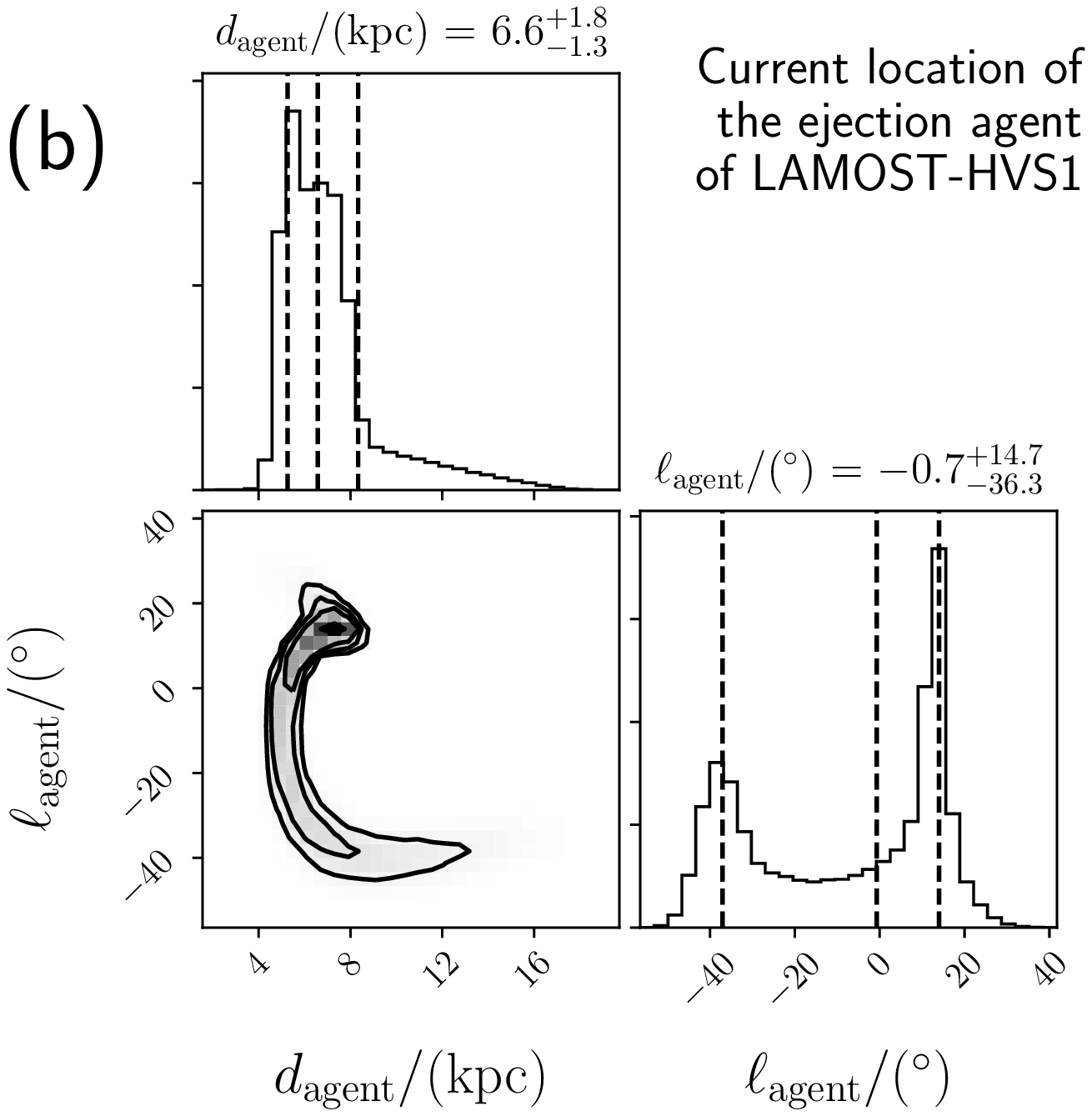}  
\end{center}
\caption{
\label{fig:triangle_distAgent_ellAgent}
The probability distribution of the current position of 
the ejection agent of \hvs\ 
(the object that ejected \hvs; 
this may be an IMBH, a VMS, or ordinary massive stars 
in the natal star cluster). 
{\bf (a)}
The distribution in the Galactic disk plane 
represented by the 103,636 Monte Carlo sample. 
{\textcolor{black}{
(Only $2\%$ of them are shown for clarity.)
}}
We note that the spatially elongated distribution  
arises from the uncertainty in the current heliocentric distance to \hvs. 
The distribution appears to have significant overlap with the Norma spiral arm.  
%(magenta curve). 
The three dashed lines indicate the 16th, 50th, and 84th percentiles of the Galactic longitude.  
The open triangles show the locations of the known young massive clusters \citep{PortegiesZwart2010ARAA}. 
{\bf (b)}
This panel shows the same information, 
but in the distance ($d_\mathrm{agent}$) and Galactic longitude ($\ell_\mathrm{agent}$) space. 
The three vertical dashed lines in the bottom right panel 
indicate the 16th, 50th, and 84th percentiles of the Galactic longitude, 
which are also shown in panel (a). 
}
\end{figure}

\subsection{Sample orbits that represent the observational uncertainty of LAMOST-HVS1's orbit} \label{sec:MC_orbits}

In reconstructing the orbit of \hvs,  we have to take into account the observational uncertainties in the current position and velocity of the star. 
Also, as mentioned in Section \ref{sec:Gaia_data},  we believe that \hvs\ is a young star ejected recently from the stellar disk.  Thus, this prior belief should be considered as well in reconstructing the orbit.  To this end, we use a Monte Carlo approach as summarized in a flow chart in Figure \ref{fig:flow_chart}. 
In the following, we describe our approach in more detail.

In order to handle the observational uncertainty, we draw 
{\textcolor{black}{
1,000,000 
}}
sets of the observed 6D position-velocity, $(d_\mathrm{spec}, \alpha, \delta,$ $\vlos, \mualpha, \mudelta)$, from the observed values of these quantities and their associated error distributions. Here, we fully take into account the correlation between $\mualpha$ and $\mudelta$, but we ignore the tiny errors on $(\alpha, \delta)$.

Each realization of 6D observable vector 
is converted to the Cartesian position $(x,y,z)$ and velocity $(v_x,v_y,v_z)$ 
at the current epoch, $t=0$. 
Then this information is used 
to compute the corresponding orbit in the past, at $-T_\mathrm{int} < t < 0$. 
Here, we adopt a conservatively long integration time $T_\mathrm{int} = 50 \Myr$, 
which is longer than our spectroscopic age estimate of $\tau = 37 \Myr$. 
We adopted this value of $T_\mathrm{int}$ 
so that systematic error on $\tau$ does not seriously affect our results.

For 
{\textcolor{black}{
628,953 of the 1,000,000 
}}
realizations (63\%), the corresponding orbit does not intersect with the disc plane at $-T_\mathrm{int} < t <0$. We discard these 
{\textcolor{black}{
improbable 
}}
%inadequate 
orbits, and we consider the remaining 
{\textcolor{black}{
371,047
}}
orbits (37\%) in the following calculation. For each of these 
{\textcolor{black}{
371,047
}}
orbits, we record the epoch $t=-T_\mathrm{flight}$ when the orbit intersects with the disk plane. 
This time is interpreted as the epoch when \hvs\ was ejected from the stellar disk, and $T_\mathrm{flight}$ corresponds to the flight time of the \hvs\ after ejection. We also record the Galactocentric position 
$(x_\mathrm{ej}, y_\mathrm{ej}, z_\mathrm{ej}=0)$ and Galactic rest frame velocity $(v_{\mathrm{ej},x}, v_{\mathrm{ej},y}, v_{\mathrm{ej},z})$ 
at the ejection epoch. In addition, we record the intrinsic ejection velocity 
\eq{
&\vector{V}_\mathrm{ej}^\mathrm{int} = 
(V_{\mathrm{ej},x}^\mathrm{int}, 
V_{\mathrm{ej},y}^\mathrm{int}, 
V_{\mathrm{ej},z}^\mathrm{int}) \nonumber \\
&= (v_{\mathrm{ej},x}, v_{\mathrm{ej},y}, v_{\mathrm{ej},z}) - 
(v_{\mathrm{circ}} \sin \phi_\mathrm{ej}, - v_{\mathrm{circ}} \cos \phi_\mathrm{ej},  0) 
\label{eq:Vejint}
}
which is measured with respect to the local circular velocity. 
The magnitude of the intrinsic ejection velocity 
$V_\mathrm{ej}^\mathrm{int} = |\vector{V}_\mathrm{ej}^\mathrm{int}|$ 
can be regarded as the impulsive velocity change that \hvs\ 
experienced when it was ejected from the stellar disk, 
with an assumption that 
it was orbiting in the Milky Way in a circular orbit before the ejection. 
Here, $v_{\mathrm{circ}}$ is the circular velocity in our potential model
evaluated at the Galactocentric cylindrical radius $R_\mathrm{ej}$; 
and we note  
$(x_\mathrm{ej},y_\mathrm{ej}) = 
(R_\mathrm{ej} \cos \phi_\mathrm{ej}, R_\mathrm{ej} \sin \phi_\mathrm{ej})$.

{\textcolor{black}{
As discussed earlier, 
\hvs\ is a massive B star 
that was probably ejected from the stellar disk. 
Thus, our initial guess 
(prior to the detailed orbital analysis)
on the probability distribution of 
the ejection location of this star, 
$(x_\mathrm{ej}, y_\mathrm{ej}, z_\mathrm{ej})$, 
is proportional to the density distribution of 
massive B-type stars in the disk. 
In order to take into account this prior distribution, 
we randomly draw 103,636 orbits from the 371,047 orbits according to a probability 
that is proportional to\footnote{
{\textcolor{black}{
A more appropriate prior should be proportional to 
$\rho (x_\mathrm{ej}, y_\mathrm{ej}, z_\mathrm{ej}) \propto 
\exp[ - R_\mathrm{ej}/ (2.5 \kpc)] 
\exp [- |z_\mathrm{ej}| / z_d]$
with $z_d$ the scale hight of B stars. 
Our assumption is equivalent to setting $z_d \to +0$, 
respecting the high concentration of B stars on the disk plane. 
}}
} 
$\Sigma (R_\mathrm{ej}) \propto \exp[ - R_\mathrm{ej}/ (2.5 \kpc)]$. 
Here, we adopt this functional form of $\Sigma (R)$ by taking into account 
the observed surface density of thin disk stars \citep{Yoshii2013}. 
Although this weighting scheme is simple, 
we find it useful 
because it does not require any knowledge on the dynamical process 
that ejected this star. 
In Section \ref{sec:result}, we will analyze these 103,636 orbits. 
}}

\subsection{Caveat}

{\textcolor{black}{
In Section \ref{sec:MC_orbits}, 
we discard 63\% of the realized Monte Carlo orbits 
because they do not cross the disk plane recently (in the last $50 \Myr$). 
We think this is mainly because 
our spectroscopic distance is not accurate enough 
(see Section \ref{sec:distance}). 
More reliable parallax data from future data releases of \Gaia\ 
would be helpful to check this view. 
If our spectroscopic distance and age are correct 
(and if the \Gaia\ proper motion is correct 
-- see Appendix\ref{sec:Gaia_quality}), 
the tension between the flight time and the stellar age 
might indicate a possibility 
that \hvs\ did not originate from the Galactic disc. 
However, 
the chemical abundance pattern 
is consistent with the disk origin of this star 
(see Table \ref{table:abundances} and Section \ref{sec:chemistry_disk}),  
so we do not pursue this unlikely possibility further in this paper. 
}}

\section{Result: Properties of the reconstructed orbit of LAMOST-HVS1}
\label{sec:result}

In Figure \ref{fig:triangle_Vej_Rej_flightTime_dist},
we plot the distributions of various quantities characterizing the ensemble of 103,636 reconstructed orbits for \hvs. This plot summarizes the most important results of this paper. Based on our orbital analysis, \hvs\ is currently located $d \simeq 13 \kpc$ away from the Sun, it was ejected from the stellar disk about $T_\mathrm{flight} \simeq 33 \Myr$ ago, its ejection location was in the inner disk ($R_\mathrm{ej} \simeq 3 \kpc$) and it was ejected with an initial intrinsic speed of  $V_\mathrm{ej}^\mathrm{int} \simeq 568 \kms$ (magnitude of the velocity vector defined in equation (\ref{eq:Vejint})). In what follows we investigate the physical implications  of these results.

\subsection{Heliocentric distance to LAMOST-HVS1} \label{sec:distance}

Figure \ref{fig:triangle_Vej_Rej_flightTime_dist} shows the probability distributions of various quantities $(V_\mathrm{ej}^\mathrm{int}, R_\mathrm{ej}, T_\mathrm{flight}, d)$ associated with the 103,636 orbits. We note that the distributions of  various quantities, especially $(R_\mathrm{ej}, T_\mathrm{flight}, d)$, are highly correlated. These correlations are, in fact, 
almost entirely explained by the distance uncertainty, 
since the observational uncertainty of the current position and velocity of \hvs\ 
is dominated by the distance uncertainty. Thus, we first investigate the heliocentric distance to \hvs\ and demonstrate that our orbital analysis actually improves our estimate on the distance to this star. 

The initial guess for the heliocentric distance was our spectroscopic distance estimate, $d_\mathrm{spec}$. 
The probability distribution of $d_\mathrm{spec}$, 
represented by the 1,000,000 Monte Carlo sample, 
is shown by the green dashed histogram in Figure \ref{fig:hist_distKin_distSpec}. 
As we can see, this distribution has a long tail toward large distance (up to $d_\mathrm{spec} \simeq 35 \kpc$).

Our orbital analysis shows that only 371,047 orbits (37\%) are 
consistent with the scenario that \hvs\ was ejected from the stellar disk recently (less than $50 \Myr$ ago). These orbits have relatively small heliocentric distance, $d \lesssim 17 \kpc$, 
as we can see from the orange histogram in Figure \ref{fig:hist_distKin_distSpec}. This preference toward small heliocentric distance is simply a consequence of requiring the flight time to be smaller than the upper limit ($50 \Myr$). %on the age of the star  

After taking into account the probability distribution of the ejection radius $R_\mathrm{ej}$ arising from the stellar density profile of the disk, the distribution of $d$ for our final sample of 103,636 orbits is even more weighted toward smaller values, as shown by the blue thick histogram in Figure \ref{fig:hist_distKin_distSpec}. 
Thus our estimate of the distance to this star 
is substantially improved by 
{\textcolor{black}{
the information on the stellar orbit,
our assumptions on the physical nature of the star, and 
properties of the stellar disk from which it was (assumed to be) ejected. 
}}
Hereafter, we will only use this posterior distribution of the distance, 
$d=13.4^{+1.7}_{-1.5} \kpc$. 
We note that this estimate happens to be close 
to the estimate by \cite{Zheng2014}.

\subsection{Current position and velocity of LAMOST-HVS1}

The orbital analysis also improves our guess on the 3D position and velocity of \hvs. 
Based on the current position of the 103,636 orbits, 
we estimate that \hvs\ is currently located at 
$(x,y,z)=(-16.2^{+0.9}_{-1.0}, -7.1^{+0.8}_{-0.9}, 7.7^{+1.0}_{-0.9}) \kpc$ 
(see Figure \ref{fig:xy}(a)(b))
with a velocity of 
$(v_x,v_y,v_z) = (-504^{+16}_{-18}$, $-131^{+7}_{-7}$, $187^{+20}_{-22}) \kms$. 
We note that the current total velocity is $v = 553^{+11}_{-9} \kms$, 
which makes this star one of the highest-velocity massive stars currently known
in the Milky Way (cf. \citealt{Zheng2014, Brown2015ARAA, Huang2017, Erkal2018, Marchetti2018, Hattori2018c, Li2018}). 
Based on these estimated phase space coordinates the  angular momentum of this star is 
$(L_x, L_y, L_z) = (
-329^{+178}_{-122},
-852^{+671}_{-824},
-1474^{+335}_{-400}) \kpc\kms$. 
We note that the value of azimuthal angular momentum $L_z$ 
is substantially different from zero, 
which suggests that \hvs\ was not ejected by the SMBH at the Galactic Center. 
Also, the prograde orbit ($L_z<0$) of this star supports the view that \hvs\ was ejected from the stellar disk. 

\subsection{Flight time $T_\mathrm{flight}$ of LAMOST-HVS1}

The orbital analysis suggests that the flight time of \hvs\ 
after it was ejected from the stellar disk is 
$T_\mathrm{flight} = 33.3^{+6.4}_{-5.0} \Myr$ 
(see Figure \ref{fig:triangle_Vej_Rej_flightTime_dist}). 
It is reassuring that the median flight time ($33.3 \Myr$) is smaller 
than the median spectroscopic age $\tau$ ($\simeq 37 \Myr$) 
(note that the spectroscopic age was not a constraint on the orbits, rather we required flight times to be less than 50~Myr). 
The proximity of the stellar age to the flight time is intriguing, suggesting that \hvs\ must have been ejected just after it was born, if we adopt our estimate of its spectroscopic age at face value. 

\subsection{Ejection location of LAMOST-HVS1}

Figure \ref{fig:xy}(c) shows the distribution of the ejection locations $(x_\mathrm{ej}, y_\mathrm{ej})$ for the ensemble of 103,636 orbits 
(small blue dots). 
The location of the Galactic Center (black $+$) and the current location of the Sun (red $\odot$) are also marked. 
The Galactocentric radius of the ejection location is 
$R_\mathrm{ej} = 2.9_{-0.9}^{+2.5} \kpc$ 
(see also Figure \ref{fig:triangle_Vej_Rej_flightTime_dist}),
where the star formation has been very active recently 
\citep{Messineo2016}.  
{\textcolor{black}{
The chemical abundance of \hvs\ is consistent 
with the view that this star was ejected from the inner stellar disk 
(see Table \ref{table:abundances} and Section \ref{sec:chemistry_disk}). 
}}

As seen from Figure \ref{fig:xy}(c), 
the distribution of $(x_\mathrm{ej}, y_\mathrm{ej})$ 
is highly elongated, 
%almost along the $x$-axis, 
and this elongated shape is almost exclusively explained by 
the uncertainty in the current heliocentric distance to \hvs. 
The ejection position $(x_\mathrm{ej}, y_\mathrm{ej})$ of a sample orbit with smaller (larger) value of $x_\mathrm{ej}$ corresponds to smaller (larger) current heliocentric distance.

As shown in Figure \ref{fig:triangle_Vej_Rej_flightTime_dist}, 
the heliocentric distance $d$ and 
the flight time $T_\mathrm{flight}$ are almost linearly correlated. 
Similarly, we found that $x_\mathrm{ej}$ and $T_\mathrm{flight}$ are almost linearly dependent.
In Figure \ref{fig:xy}(c), 
an orbit with $x_\mathrm{ej} \simeq -4.7 \kpc$ 
has $T_\mathrm{flight} \simeq 16.7 \Myr$, 
while an orbit with $x_\mathrm{ej} \simeq 11 \kpc$ 
has $T_\mathrm{flight} \simeq 50 \Myr$ (the maximum allowed flight time).  
There are no sampled orbits with $x_\mathrm{ej}> 11.4 \kpc$ 
due to our limit on $T_\mathrm{flight}$.

The Galactocentric radius of the ejection location 
$R_\mathrm{ej} = \sqrt{x_\mathrm{ej}^2 + y_\mathrm{ej}^2}$ 
attains its minimum when $d \simeq 12 \kpc$ 
(and $x_\mathrm{ej} \simeq 0 \kpc$). 
The value of $R_\mathrm{ej}$ becomes larger for $d < 12 \kpc$ 
and $d > 12 \kpc$ 
since $|x_\mathrm{ej}|$ becomes larger 
(while $y_\mathrm{ej}$ varies only mildly as a function of $d$). 
This is why we see a `L'-shaped correlation between 
$R_\mathrm{ej}$ and $d$ 
in Figure \ref{fig:triangle_Vej_Rej_flightTime_dist}. 

\subsection{Intrinsic ejection velocity of LAMOST-HVS1}

Given the current position and velocity of \hvs, 
the $x$-component of its velocity $v_{\mathrm{ej},x}$ just after the ejection 
must have been a large negative value. 
If \hvs\ had a circular orbit before the ejection, 
the intrinsic ejection velocity $V_\mathrm{ej}^\mathrm{int}$ 
is smallest 
when the velocity vector of the circular orbit is most closely aligned with 
the velocity just after the ejection, 
$(v_{\mathrm{ej},x}, v_{\mathrm{ej},y}, v_{\mathrm{ej},z})$. 
This minimum value of $V_\mathrm{ej}^\mathrm{int}$ is attained when 
$x_\mathrm{ej} \simeq 1.5 \kpc$ and 
$R_\mathrm{ej} \simeq 2.2 \kpc$. 
This is the reason for the `L'-shaped correlation between 
$R_\mathrm{ej}$ and $V_\mathrm{ej}^\mathrm{int}$ 
in Figure \ref{fig:triangle_Vej_Rej_flightTime_dist}.

It is important to note that even in the extreme case 
when the intrinsic ejection velocity 
$\vector{V}_\mathrm{ej}^\mathrm{int}$ is most closely aligned to the circular orbit rotating with the circular velocity of the disk at that radius, the magnitude of 
$V_\mathrm{ej}^\mathrm{int}$ is as large as $530 \kms$. 
The intrinsic ejection velocity of 
$V_\mathrm{ej}^\mathrm{int} \simeq 568^{+19}_{-17} \kms$ 
is much larger than those of other massive B-type runaway stars
(see Sections \ref{sec:comparison} and \ref{sec:other_fast_stars}).

\subsubsection{Comparison with nearby massive runaway stars in \cite{SN2011}}
\label{sec:comparison}

In order to demonstrate that 
the massive runaway star \hvs\ is 
exceptional in its large 
$V_\mathrm{ej}^\mathrm{int}$, 
we also analyzed known runaway B-type massive stars in \cite{SN2011}. 
\cite{SN2011} analyzed the orbits of 96 stars that they classified as main-sequence stars. 
We found that \Gaia\ data is available for all of these stars. 
For the 46 stars in this sample with high quality parallax data 
($\varpi/\sigma_\varpi>5$), 
we derived the probability distribution of $V_\mathrm{ej}^\mathrm{int}$, 
and derived the 16th, 50th, and 84th percentile values of $V_\mathrm{ej}^\mathrm{int}$. 
We found that only three stars 
(\runawayI, \runawayII, \runawayIII)
have $V_\mathrm{ej}^\mathrm{int}$ that is consistent with $300 \kms$ or larger 
(i.e., the 84th percentile value of $V_\mathrm{ej}^\mathrm{int}$ is larger than $300\kms$). 
Figure \ref{fig:hist_vej} shows the probability distribution 
of $V_\mathrm{ej}^\mathrm{int}$ for these three stars. 
Also, we show the combined probability distribution of 
$V_\mathrm{ej}^\mathrm{int}$ 
for the remaining 43 high-quality sample. 
From this figure, we see that $V_\mathrm{ej}^\mathrm{int}$ of \hvs\ 
is much larger than that of any other nearby massive B-type runaway stars with high-quality parallax.

We note that the sample in \cite{SN2011} includes 
some stars that are several $\kpc$ away from the Galactic disk 
according to their spectroscopic distance. 
Since our cut on the fractional error in parallax ($\varpi/\sigma_\varpi$) preferentially excludes distant stars,  some of these distant runaway stars might have a very large  $V_\mathrm{ej}^\mathrm{int}$. 
However, since the main focus of this paper is \hvs, 
we do not pursue this point in this paper.

\subsubsection{Comparison with other fast-moving stars}
\label{sec:other_fast_stars}

Prior to \Gaia\ DR2, more than 20 hypervelocity star candidates 
have been studied \citep{Brown2015ARAA,Huang2017}. 
The re-analysis of these stars after \Gaia\ DR2 
shows that some fraction of them are probably runaway stars ejected 
from the Galactic disk \citep{Erkal2018,Brown2018,Irrgang2018}. 
Most of these stars are late B-type or A-type main-sequence stars 
with $M \lesssim 4 ~\msun$ \citep{Brown2014}, 
and only a few of them are as massive as \hvs.

For example, \hd\ is a 11-$\msun$ runaway star 
that was ejected from the outskirt of the Galactic stellar disk \citep{Heber2008}. 
Due to its large current velocity, 
\hd\ is classified as a hyper-runaway star. 
However, we have confirmed based on the \Gaia-DR2 proper motion data that 
the intrinsic ejection velocity of \hd\ is 
$V_\mathrm{ej}^\mathrm{int} \simeq 400 \kms$, %$=368^{+22}_{-7} \kms$ [with prior], 
which is not as large as that of \hvs.

Also, an early B-type star \hvstwo\ \citep{Huang2017} 
is a good candidate for a massive hyper-runaway star. 
Our (unpublished) preliminary analysis suggests that this star 
has a high probability of having been ejected from the Galactic stellar disk, consistent with the study by \cite{Erkal2018}. 
However, due to the lack of high-resolution spectrum of this star, 
we cannot rule out the possibility that this star is 
a low-mass blue star.

HIP 60350 is another example of high-velocity runaway star, 
but this star is only $4.9\msun$ \citep{Irrgang2010}
and not as massive as \hvs. 

\cite{Li2018} discovered 
a hyper-runaway candidate (\hvsfour),  
but this star is only $\sim4 ~\msun$ if it is a main-sequence star. 
Also, there is a possibility that this star is a low-mass BHB star due to the low-resolution spectra currently available for this star.

HVS3 (HE 0437-5439) %(a well-known blue star with extremely large velocity) 
is as massive as \hvs, 
but it may have originated from the Large Magellanic Cloud 
\citep{edelman_etal_2005,Gualandris2007,Przybilla2008,
Bonanos2008,Brown2015ARAA,Erkal2018,Irrgang2018}.

Based on these considerations,
we argue that \hvs\ is the only well-confirmed massive hyper-runaway star 
ejected from the Galactic stellar disk 
with extremely large intrinsic ejection velocity of 
$V_\mathrm{ej}^\mathrm{int} \sim 600 \kms$.

\subsection{Current location of the ejection agent of LAMOST-HVS1}

Theoretical investigations have suggested 
some possible mechanisms to accelerate disk stars 
to produce runaway stars with extreme velocity. 
In all of these mechanisms, some nearby object 
---including intermediate mass black hole (IMBH) or very massive star---
plays an important dynamical role in ejecting a disk star. 
To date there is no conclusive evidence that an IMBH exists in the Milky Way 
(although there have been claims of objects close to the central SMBH; 
e.g. \citealt{Oka2017}). 
While a few very massive stars have been discovered 
(e.g. the initial mass of Pistol star is estimated to be 200-250 $\msun$; \citealt{Figer1998}), 
there is no consensus on the numbers of such stars that exist 
in the Galaxy and under what specific conditions they form.  
Thus, there are compelling reasons 
to investigate where the `ejection agent' of \hvs\ 
(the object that took part in the ejection of \hvs) 
is currently located.

Since the ejection agent was located at $(x_\mathrm{ej}, y_\mathrm{ej}, 0)$ 
at $t=-T_\mathrm{flight}$,
we can derive its current position 
by assuming the velocity at $t=-T_\mathrm{flight}$  
and integrating the orbit in our model potential 
(Section \ref{sec:model_potential}) until $t=0$. 
For each of the 103,636 sample orbits, we assume that 
the velocity of the ejection agent at $t=-T_\mathrm{flight}$ 
is given by 
$(v_x,v_y,v_z) = 
(v_{\mathrm{circ}} \sin \phi_\mathrm{ej} + v_{\mathrm{pec},x}, 
- v_{\mathrm{circ}} \cos \phi_\mathrm{ej} + v_{\mathrm{pec},y}, 0)$. 
Here, the peculiar velocity (deviation from the circular velocity) 
$v_{\mathrm{pec},x}$ and $v_{\mathrm{pec},y}$ 
are randomly drawn from a 2D isotropic Gaussian distribution with a dispersion of $20 \kms$ (typical of the velocity dispersion of young stars in the Milky Way). 
We note that the orbit of the ejection agent is assumed to be confined 
to the Galactic disk plane.

Figure \ref{fig:triangle_distAgent_ellAgent}(a) shows (in blue dots) the possible current location of the ejection agent in the Galactic plane 
represented by the 103,636 Monte Carlo samples. 
To guide the eye, we draw three dashed lines 
corresponding to the 16th, 50th, and 84th percentile values of  
probable Galactic longitude of the ejection agent. 
Also, the curved solid lines of various colors represent the current positions of the known prominent spiral arms: Norma (magenta), Perseus (yellow), Sagittarius-Carina (green) and Scutum-Cruz (blue) \citep{Vallee2008}. 
From this figure, we see that the probability distribution of the current location of the ejection agent shows an elongated shape, 
reflecting the elongated distribution of $(x_\mathrm{ej}, y_\mathrm{ej})$. 
We also note that this elongated distribution is bent 
at a dashed line of $\ell = 14.0^\circ$. 
This structure is due to the differential rotation of the disk. 
Since the angular velocity of disk stars is larger at smaller $R$, 
those ejection agents with smaller $R_\mathrm{ej}$ 
rotate around the Milky Way at larger angular velocity. 
We also note that the Monte Carlo points (blue dots)
are nearly aligned with the Norma spiral arm (magenta curve) 
at $-37.1^\circ < \ell < 14.0^\circ$, 
which corresponds to the central 68\% of the distribution.
This is consistent with the idea that young massive star clusters, where massive stars are thought to form, are expected to be located in or close to spiral arms.

Figure \ref{fig:triangle_distAgent_ellAgent}(b) shows 
the probability distributions for the current location of the ejection agent in the $(d_\mathrm{agent}, \ell_\mathrm{agent})$-plane. 
We see that the Galactic longitude $\ell_\mathrm{agent}$ has a wide distribution 
with two prominent peaks. 
These peaks happen to be located near the 
84th and 16th percentiles of $\ell_\mathrm{agent}$, 
at $\ell_\mathrm{agent} = 14.0^\circ$ and $-37.1^\circ$. 
One of the peaks at $\ell_\mathrm{agent} = 14.0^\circ$ 
approximately corresponds to the bending point of 
$(x_\mathrm{agent}, y_\mathrm{agent})$. 
We found that $20\%$ of the orbits in the distribution are enclosed 
at $11.8^\circ<\ell_\mathrm{agent}<15.4^\circ$, 
and the corresponding heliocentric distance at this longitude range  
is $7.2^{+0.6}_{-0.7} \kpc$. 
Another peak at $\ell_\mathrm{agent} = -37.1^\circ$ 
corresponds to the tail of the distribution at $y_\mathrm{agent}<-4 \kpc$, 
whose distribution is nearly aligned with the line of $\ell_\mathrm{agent} = -37.1^\circ$.

\subsubsection{LAMOST-HVS1 indicates an undiscovered young massive cluster (YMC)?} \label{sec:undiscovered_YMC}

In the above calculation,
we have implicitly assumed that the 
ejection agent still survives until today. 
This is probably the case if the ejection agent is an IMBH, 
but not the case if it is a reasonably massive star ($\gtrsim 30~\msun$) 
due to the supernova explosion. 
Even if the ejection agent does not survive until today, 
our calculation is still useful, 
since it shows the current location of the natal star cluster of \hvs.

If the natal cluster of \hvs\ is as massive as $10^4 ~\msun$ 
(which is a reasonable assumption as we will describe in 
Sections \ref{sec:mechanism} and \ref{sec:rate}) 
located at $R \sim 3\kpc$, 
the cluster can survive for $\sim 1 \Gyr$ 
according to equation (19) of \cite{PortegiesZwart2010ARAA}. 
Even if the cluster mass is $\sim 10^3 ~\msun$, it survives for $\sim 200 \Myr$. 
This means that the natal cluster may well survive until today. 
As shown in Figure \ref{fig:triangle_distAgent_ellAgent}(a), 
only three young massive clusters (YMCs) (RSGC1, RSGC2, and RSGC3) 
at around $(x,y)=(-2.8, 2.5) \kpc$ 
are marginally consistent with the region 
where the natal cluster of \hvs\ can be located. 
However, since RSGC1-3 are only 12-18 $\Myr$ old \citep{PortegiesZwart2010ARAA}, 
these clusters did not exist when \hvs\ was ejected (about $33 \Myr$ ago). 
{\textcolor{black}{
Moreover, the sub-Solar abundance of O, Mg, and Si of RSGC1-3 \citep{Davies2009b,Origlia2016}
is also inconsistent with the super-Solar abundance of these elements for \hvs. 
}}
Therefore, none of the known YMCs could have ejected \hvs. 
This result indicates that 
the natal cluster of \hvs\ 
is an undiscovered YMC probably located along the Norma spiral arm 
(that may be heavily dust obscured). 
{\textcolor{black}{
The chemical abundance pattern of \hvs\ 
(see Table \ref{table:abundances}) 
might be useful to identify its natal YMC 
or its sibling stars 
that were born in the same cluster. 
Such a `chemical-tagging' 
\citep{Freeman2002ARAA,Hawkins2018} 
may help our understanding 
on the origin of hyper-runaway stars. 
}}

\subsection{Chemistry of LAMOST-HVS1} 
\label{sec:chemistry_disk}

{\textcolor{black}{
As we have shown, 
the kinematic data of \hvs\ 
indicate that it was ejected from the inner stellar disk. 
Here we show that its chemical data 
are consistent with this view. 
}}

{\textcolor{black}{
Young stars in the Galactic disk (e.g., classical Cepheids and OB stars)
are known to have a negative abundance gradient such that 
the inner stellar disk is more metal-rich than in the outer disk
\citep{Carigi2005,Esteban2005,Cescutti2007}. 
As seen in Table \ref{table:abundances}, 
the abundance of Mg, C, N, and O 
of \hvs\ is systematically higher than that of 
the mean abundance 
$\langle \mathrm{A}_\mathrm{B}^{8 \kpc} \rangle$ 
of B stars at $R=8 \kpc$, 
while it is consistent with 
$\langle \mathrm{A}_\mathrm{B}^{3 \kpc} \rangle$ 
at $R=3 \kpc$. 
The similarity in the chemical abundance pattern 
between \hvs\ and B stars at $R=3 \kpc$
supports our estimation that 
the ejection radius of \hvs\ was at $R_\mathrm{ej} = 2.9_{-0.9}^{+2.5} \kpc$. 
}}

{\textcolor{black}{
The Si abundance of \hvs, in contrast, 
is appreciably higher than that of typical B stars at $R=3 \kpc$.  
This excess of Si is intriguing, 
because another $\alpha$ element Mg 
(whose origin is similar to that of Si)
does not show such an excess. 
In order to see if the excess of Si is anomalous,
we checked the relative abundance of Si and Mg, $\mathrm{A(Si)- A(Mg)}$, 
for Cepheids across the Galactic disk.
To this end, 
we used 231 Cepheids in \cite{Luck2011} 
for which both [Si/H] and [Mg/H] are available. 
We used [Si/H] and [Mg/H] for these stars and 
\cite{Grevesse1996} Solar abundance 
to derive $\mathrm{A(Si)- A(Mg)}$. 
We did not see a clear trend of $\mathrm{A(Si)- A(Mg)}$
as a function of $R$, and 
$\mathrm{A(Si)- A(Mg)}$ has the mean value of $-0.01$ 
and the standard deviation of $0.13$. 
Among these 231 Cepheids, 
we found 10 stars that show $\mathrm{A(Si)- A(Mg)}>0.18$, 
and 3 stars with $\mathrm{A(Si)- A(Mg)}>0.30$. 
Thus, the observed $\mathrm{A(Si)- A(Mg)}=0.18\pm0.10$ for \hvs\ 
may not be exceptionally high.\footnote{ 
{\textcolor{black}{
\cite{McEvoy2017} performed 
a non-LTE abundance analysis for 39 B-type runaway stars 
and found that 31 of them show $\mathrm{A(Si)- A(Mg)} > 0$. 
This result indicates that 
the overabundance of Si over Mg may not be too peculiar for runaway stars. 
}}
}
The excess of Si for \hvs\ might be helpful 
to identify its natal star cluster or its sibling stars 
(see Section \ref{sec:undiscovered_YMC}). 
}}

\section{Ejection mechanism of LAMOST-HVS1} 
\label{sec:mechanism}

As reviewed in Section \ref{sec:intro}, 
there are two possible mechanisms, BEM and DEM, 
to eject a runaway star. Here we investigate which of these mechanisms  could be responsible for the hyper-runaway star \hvs. 

\subsection{Binary ejection mechanism (BEM)}
 
In BEM, 
the intrinsic ejection velocity of 
a runaway star that is as massive as \hvs\ 
($M \simeq 8 M_\odot$) 
is expected to be smaller than $\sim 400 \kms$ 
(see Section \ref{sec:intro}). 
This upper limit is significantly smaller than 
the ejection velocity 
$V_\mathrm{ej}^\mathrm{int} \simeq$ $568^{+19}_{-17}$ $\kms$ 
of \hvs. 
Thus, BEM is disfavored by our estimate of $V_\mathrm{ej}^\mathrm{int}$.

\subsection{Dynamical ejection mechanism (DEM)}

In DEM, the intrinsic ejection velocity can be as large as 
$V_\mathrm{ej}^\mathrm{int} \sim 10^3 \kms$, 
depending on the situation 
(see references in Section \ref{sec:intro}). 
Thus our estimate of the ejection velocity for \hvs\ 
is compatible with DEM. 
Here we further investigate the required situation 
for realizing the ejection velocity of \hvs.

\subsubsection{3-body interaction including an IMBH}

Hills mechanism is a dynamical mechanism 
in which a massive compact object (e.g., a black hole) 
captures one star in a stellar binary that passes nearby 
and ejects the other star with a large velocity 
\citep{Hills1988,YuTremaine2002}. 
\cite{Gualandris2007} and \cite{Gvaramadze2008} investigated 
how an IMBH ejects an 8-$\msun$ main-sequence star 
with Hills mechanism. 
Their works are useful for us since \hvs\ is also $\simeq 8 ~\msun$. 
They estimated the probability that an 8-$\msun$ star attains 
$V_\mathrm{ej}^\mathrm{int} > 500 \kms$ during this process. 
They found that this probability is larger  
(i) if the IMBH is more massive; 
(ii) if the binary companion of the 8-$\msun$ star is more massive; 
or 
(iii) if the initial orbital separation $a$ of the binary 
is smaller (more compact binary). 
For example, 
when an equal-mass binary consisting of two 8-$\msun$ stars
interacts with an IMBH with $M_\mathrm{BH} = 10^2 ~\msun$, 
there is $\sim 1\%$ probability that 
one of the 8-$\msun$ star attains 
larger ejection velocity than \hvs. 
This probability increases to $\sim 15\%$ and $\sim 30\%$, 
if $M_\mathrm{BH}=10^3 ~\msun$ and $10^4 ~\msun$, respectively. 
Thus, this mechanism can explain the ejection of \hvs.

If \hvs\ was ejected by an IMBH, 
then we expect that this IMBH is currently located at 
the region shown in Figure \ref{fig:triangle_distAgent_ellAgent}. 
If this hypothetical IMBH captures a nearby star to form a close binary,
it may emit ultra-luminous X-rays ($>10^{39} \;\mathrm{erg\;s^{-1}}$) 
by accreting material 
from the binary companion \citep{Hopman2004}. 
There are no observational signatures of such ultra-luminous X-ray sources 
in that region (or elsewhere in the Milky Way). 
This may be explained 
(i) if the hypothetical IMBH currently does not have a close binary companion; 
(ii) if the close binary companion has stopped transferring material to the IMBH; 
or 
(iii) if the X-ray source is highly obscured by dust. 
Even if the hypothetical IMBH does not have a close binary companion, 
it could emit (lower-luminosity) X-rays 
($<10^{39} \;\mathrm{erg\;s^{-1}}$) 
from an accretion disk formed from ambient molecular gas in the natal cloud. 
Such X-rays might have been already observed 
as an unclassified hard X-ray source on the Galactic disk plane 
\citep{Fornasini2017,Krivonos2017,Oh2018}. 
In this regard, it is interesting to note that \cite{Fornasini2017} published some ill-characterized X-ray sources in the Norma spiral arm region (such as the source named `NNR 28' in \citealt{Fornasini2017}).

\subsubsection{3-body interaction including a very massive star (50-100 $M_\odot$ or more)}
\label{sec:3bodyVMS}

Hills mechanism is effective not only around an IMBH, 
but also around other massive compact objects 
such as very massive stars (VMSs) 
with a mass $M_\mathrm{VMS}=$ 50-100 $M_\odot$ or more. 
Such stars may have formed as a result of runaway mergers of less massive stars 
in dense star clusters \citep{Miller2002,PortegiesZwart2004}. 
The Pistol star in the Arches cluster,
whose initial mass is estimated to be 200-250 $~\msun$ \citep{Figer1998},
could be an example of such a star.

\cite{Gvaramadze2009} showed that 
when a compact stellar binary 
with the initial separation of $a=0.15$ AU 
consisting of main-sequence stars 
with mass of $(8~\msun, 40~\msun)$ 
interacts with a 100-$\msun$ VMS, 
the Hills mechanism ejects
the 8-$\msun$ star with a velocity of 
$V_\mathrm{ej}^\mathrm{int} > 600 \kms$ 
(similar to $V_\mathrm{ej}^\mathrm{int}$ of \hvs) 
with a probability of 1\%. 
This probability is increased if $M_\mathrm{VMS}$ is larger, 
and is decreased if $a$ is larger, 
which are consistent with the similar results in 
\cite{Gualandris2007}. 
Thus, the ejection of \hvs\ can be explained with this mechanism.

\subsubsection{4-body interaction including a $\sim 30$-$M_\odot$ star}
\label{sec:4body883232}

According to Table 1 of \cite{Leonard1991},
when two equal-mass binaries consisting of main-sequence stars with mass of $(m,m)$ and $(4m,4m)$ collide, 
the least massive star ($m$) can attain 
an ejection velocity up to 
$0.721 v_{\mathrm{esc},*} {(4m)}$, 
where $v_{\mathrm{esc},*} {(4m)}$ 
is the stellar surface escape velocity of the most massive star ($4m$). 
If we set $m=8 ~\msun$, the corresponding ejection velocity 
of an 8-$\msun$ star is $770 \kms$ ($= 0.721 \times 1069 \kms$). 
(Here we assume the radius of a 32-$\msun$ star 
to be $10.7 ~R_\odot$ following 
a 3-Myr old stellar model from \citealt{2012A&A...537A.146E}.)
This maximum ejection velocity is larger than 
our estimate of $V_\mathrm{ej}^\mathrm{int}$ for \hvs, 
so this mechanism can explain the ejection of this star.

\subsubsection{4-body interaction with stars with 8-$M_\odot$ or less}

\cite{Leonard1991} claimed that 
when two identical equal-mass binaries 
consisting of main-sequence stars 
with mass of $(m,m)$ and $(m,m)$ collide, 
one of the stars with mass $m$ can attain 
an ejection velocity up to 
$0.5 v_{\mathrm{esc},*} {(m)}$. 
This upper limit remains the same or decreases 
if we replace some of the non-ejected stars with less massive stars.  
If we set $m=8 ~\msun$, the corresponding ejection velocity 
of an 8-$\msun$ star is $477 \kms$ ($= 0.5 \times 954 \kms$). 
(Here we assume the radius of an 8-$\msun$ star 
to be $3.36 ~R_\odot$ following \citealt{2012A&A...537A.146E}.)
This maximum ejection velocity is smaller than 
our estimate of $V_\mathrm{ej}^\mathrm{int}$ for \hvs.
Therefore, the ejection of \hvs\ 
cannot be explained by dynamical interaction of stars 
if all of the interacting stars are as massive as or 
less massive than \hvs.

\section{Ejection frequency}
\label{sec:rate}

An 8-$\msun$ star, like \hvs, can survive for $\sim 30 \Myr$. 
Thus, observable hyper-runaway stars more massive than $\sim 8 ~\msun$ 
must have been ejected in the last $\sim 30 \Myr$ or so. 
Currently, there is only one star, \hvs, that is confirmed (with high confidence)
to be a massive  hyper-runaway star with $V_\mathrm{ej}^\mathrm{int} \gtrsim 600 \kms$.
(We note that \hd\ is also a massive hyper-runaway star, 
but with $V_\mathrm{ej}^\mathrm{int} \simeq 400 \kms$.) 
Thus, we estimate that the ejection rate of 
massive ($\geq 8 ~\msun$) hyper-runaway stars 
with $V_\mathrm{ej}^\mathrm{int} \gtrsim 600 \kms$ 
from the entire stellar disk 
of the Milky Way is at least $\sim 1$ per $30 \Myr$.

In Section \ref{sec:mechanism}, we argued that the large intrinsic ejection velocity of \hvs\ is consistent with three channels of dynamical ejection. 
Here we estimate the expected number of massive hyper-runaway stars with $V_\mathrm{ej}^\mathrm{int} \gtrsim 600 \kms$ ejected in the last 30 Myr for these mechanisms under some optimistic but reasonable assumptions. These theoretical ejection rates would be helpful to determine whether any of the channels can naturally explain the observed ejection rate.

In Section \ref{sec:review_YMC}, 
we explain some common assumptions. 
In Sections \ref{sec:gamma_massive}, 
\ref{sec:gamma_IMBH}, and 
\ref{sec:gamma_VMS}, 
we will discuss the theoretical ejection rates associated with 
(1) ordinary massive stars, (2) an IMBH, and (3) a VMS, respectively. 
A busy reader may skip to Section \ref{sec:summary_ejection}, 
where we summarize these ejection rates.

\subsection{Assumptions on young massive clusters (YMCs)}
\label{sec:review_YMC}

Since all of the possible mechanisms 
to eject a hyper-runaway star with $M=8~\msun$ 
are associated with close encounter of the 8-$\msun$ star with more massive objects 
(e.g an IMBH or VMS) and since these objects are thought to preferentially form 
in regions of very high stellar density \citep{Gvaramadze2008,Gvaramadze2009}, 
we assume, as others have done, that these massive objects form rapidly  
in YMCs. 
Since these mechanisms also operate most efficiently in YMCs 
we assume them to be the sites of massive hyper-runaway ejection. 
In this Section, we briefly summarize some of the assumptions and concepts 
that we use in our calculations to estimate the ejection rates.

\subsubsection{Stars in YMCs}
\label{sec:IMF}

There have been extensive efforts to constrain the initial mass function (IMF) 
of stars in YMCs, but due to the limited range of detectable mass, 
it is currently difficult to determine whether their IMFs differ significantly from 
standard IMF models \citep{PortegiesZwart2010ARAA}. 
Throughout this Section, we assume that the stars in a star cluster 
follow the Salpeter initial mass function \citep{Salpeter1955} 
for stellar masses in the range $0.2 ~\msun < M < 60 ~\msun$. 
Under this assumption, the mean stellar mass is $\langle M \rangle = 0.667 ~\msun$ 
and each cluster contains $N_* = 1.50 (M_\mathrm{cl}/\msun)$ stars 
when the cluster is formed.  Star clusters with mass 
$M_\mathrm{cl} < 10^4 ~\msun$ contain less than 10 stars with $M>30 ~\msun$, 
so we will mainly focus on massive clusters with $M_\mathrm{cl}\gtrsim 10^4 ~\msun$.  
For a Salpeter IMF, the number of massive stars with 
$20~\msun <M<60~\msun$ and $7~\msun <M<11~\msun$ 
in a cluster with mass $M_\mathrm{cl}$ are given by 
\eq{
N^{*}_{(20{\text -}60)} =2.31 \times 10^{-3} M_\mathrm{cl}/\msun
}
and 
\eq{
N^{*}_{(7{\text -}11)} =5.64 \times 10^{-3} M_\mathrm{cl}/\msun, %7-11msun
}
respectively. 
In the following arguments,  for simplicity,  we treat each star with (20-60)-$\msun$ as if it is a 32-$\msun$ star. This is justified by the fact that the mean mass 
for these stars is $\simeq 32~\msun$. Similarly,  we treat each star with (7-11)-$\msun$ as if it is a 8-$\msun$ star. This is motivated by the fact that we are interested in hyper-runaway stars with $M\simeq 8 ~\msun$.

For massive stars with $M> 7 ~\msun$, we assume that the binary fraction is $80\%$, which is motivated by  the observed multiplicity fraction of $\gtrsim (60{\text -}80)\%$ for high-mass stars \citep{Chini2012,Duchene2013ARAA}. 
Also, 
the orbital semi-major axis of each massive binary
(a binary containing massive stars with $M>7 \msun$) 
is assumed to be $a=30 ~R_\odot (\simeq 0.14 \AU)$, 
which roughly corresponds to the 
peak (but not median) of the observed distribution of $a$ 
for massive binaries (see fig 2 of \citealt{Duchene2013ARAA}).

We assume primordial mass segregation in YMCs. 
This means that 
massive stars are already located at the core region when a YMC is formed. 
This kind of primordial mass segregation is discussed by \citet{Gvaramadze2009}, 
but we note that similar mass segregation may be achieved 
by dynamical friction \citep{Gvaramadze2008}. 
Also, 
motivated by numerical simulations \citep{PortegiesZwart2010ARAA}, 
we assume that 
the core radius stays small ($r_\mathrm{c} \simeq 0.1 \pc$) 
for only $\sim 6 \Myr$; 
and after that the core region expands 
due to the mass loss from massive stars or supernova explosion. 
The timescale for core expansion 
is important in our calculations, 
since the ejection rate of massive hyper-runaway stars depends on 
the number density of massive stars at the core. 
For reference, the main-sequence lifetime of 
8-, 30-, 60-$\msun$ star are $30 \Myr$, $6\Myr$, and $3 \Myr$,
respectively \citep{2012A&A...537A.146E}.

\subsubsection{Cluster mass function} \label{sec:clusterMF}

Since we are interested in the number of massive hyper-runaway stars 
ejected from the Milky Way 
in the last $30\Myr$, 
it is important to estimate the number of YMCs 
(the assumed ejection sites) formed in the last $30 \Myr$. 
Following Section 2.4.2 of \cite{PortegiesZwart2010ARAA}, 
we assume that the star formation rate in the Solar neighborhood is 
$\sim 5\times10^3 ~\msun \Myr^{-1} \kpc^{-2}$ 
and that $\sim 10\%$ of the mass is contained in bound star clusters (including YMCs)
that do not expand and dissolve. 
Then the total mass of bound star clusters 
formed within $R<10 \kpc$ from the Galactic Center in the last $30 \Myr$ is 
$M_\mathrm{tot}= 4.71 \times 10^6 ~\msun$. We assume that the mass function of star clusters formed in the last $30\Myr$ is given by a Schechter function 
with a power-law index of $(-2)$ \citep{Lada2003ARAA} 
and scale mass of $M_\mathrm{Sch}=2\times10^5 ~\msun$ 
\citep{PortegiesZwart2010ARAA}. 
If the cluster mass range is 
$50 ~\msun < M_\mathrm{cl} < 4\times 10^5 ~\msun$, 
the mass function is given by  
\eq{\label{eq:dN_dM}
\frac{\mathrm{d}N_\mathrm{cl}}{\mathrm{d}M_\mathrm{cl}}
= \frac{M_\mathrm{tot}}{{\textcolor{black}{7.6682}}} M_\mathrm{cl}^{-2} 
\exp \left[- \left( \frac{M_\mathrm{cl}}{M_\mathrm{Sch}} \right) \right].
}
In this case,  
we have 546, 49 and 2 star clusters 
with mass of 
$M_\mathrm{cl}=(10^3{\text -}10^4)~\msun$,
$M_\mathrm{cl}=(10^4{\text -}10^5)~\msun$, and 
$M_\mathrm{cl}=(10^5{\text -}10^{5.6})~\msun$,  
respectively (see also \citealt{Gvaramadze2008,Gvaramadze2009}). 
This estimate is consistent with recent observations of YMCs more massive than 
$10^4 ~\msun$. 
According to table 2 and fig 3 of \cite{PortegiesZwart2010ARAA},
there are 12 YMCs younger than $\sim 20 \Myr$ 
and almost all of these YMCs are distributed in the near-side of the disk. 
Based on the spatial distribution, 
we estimate that the completeness of this YMC sample 
with $M_\mathrm{cl} \gtrsim 10^4 ~\msun$ is 
$f_\mathrm{completeness}\sim {\text{0.25-0.50}}$. 
Then the number of YMCs with $M_\mathrm{cl} \gtrsim 10^4 ~\msun$
formed in the last $30 \Myr$ is 
$N_\mathrm{cl} \sim (30 \Myr / 20 \Myr) \times 12 / f_\mathrm{completeness} 
= ({\text{36-72}})$.

\subsection{Ejection frequency by massive stars with $\simeq 30~\msun$}
\label{sec:gamma_massive}

Here we investigate the ejection frequency of 
early B-type main-sequence stars with $\simeq 8 ~\msun$  
by 4-body interaction with more massive stars 
($\sim 30 ~\msun$; see Section \ref{sec:4body883232}) 
in the center of YMCs.

To make our calculations tractable, 
we specifically consider interaction between 
$(8~\msun, 8~\msun)$- and 
$(32~\msun, 32~\msun)$-binaries (see Section \ref{sec:IMF}).
In addition, we assume, for simplicity, 
that all binaries are equal-mass binaries. 
With these assumptions as well as the assumptions in Section \ref{sec:IMF}, 
the number of binaries in the core of a YMC with a mass $M_\mathrm{cl}$
can be given by 
$N_{(32,32)}^\mathrm{bin} = 0.924 \times 10^{-3} M_\mathrm{cl}/\msun$ 
for binaries with mass $(32 ~\msun, 32 ~\msun)$ 
and 
$N_{(8,8)}^\mathrm{bin} = 2.26 \times 10^{-3} M_\mathrm{cl}/\msun$ 
for binaries with mass $(8 ~\msun, 8 ~\msun)$.

The rate of close encounters between 
these binaries is given by 
\eq{ \label{eq:Gamma_binbin_encounter}
\Gamma^\mathrm{encounter}_{\text{bin-bin}} \simeq 
\left( \frac{4\pi}{3} r_\mathrm{c}^3 \right)^{-1} 
N^\mathrm{bin}_{(8,8)} 
N^\mathrm{bin}_{(32,32)} 
\sigma_{\text{bin-bin}} V_\mathrm{rel},
}
where $\sigma_{\text{bin-bin}}$ 
is the cross section of binary-binary interaction 
and $V_\mathrm{rel}$ is the relative velocity of binaries. 
Unlike \cite{Leonard1989} or \cite{Gvaramadze2008}, 
we do not multiply by a factor $1/2$ in evaluating 
$\Gamma^\mathrm{encounter}_{\text{bin-bin}}$ 
in equation (\ref{eq:Gamma_binbin_encounter}),
since we distinguish the two types of binaries. 
If we assume that only 1 \% of close encounters results in 
an ejection of a massive hyper-runaway star 
with $V_\mathrm{ej}^\mathrm{int} \gtrsim 600 \kms$ 
(cf. \citealt{Leonard1991}),
we obtain the ejection rate for massive hyper-runaway stars
\eq{ \label{eq:Gamma_binbin}
\Gamma_{\text{bin-bin}} =  0.01 \times \Gamma^\mathrm{encounter}_{\text{bin-bin}} .
}

We substitute $\sigma_\mathrm{bin,bin}$ with 
the cross section between a $(8~\msun,8~\msun)$- 
and $(32~\msun,32~\msun)$-binaries. 
This procedure is motivated by the 
simulations by \cite{Leonard1991}. 
We assume that the orbital semi-major axis in each binary is 
$a=30 ~R_\odot$. 
Also, following \cite{Leonard1989},
we set the pericenter distance between these binaries to be $r_\mathrm{peri}=a$. 
These assumptions lead to a cross section of 
\eq{ \label{eq:sigma_binbin}
\sigma_\mathrm{bin,bin} \simeq 
\frac{2\pi G (m^\mathrm{bin}_1 + m^\mathrm{bin}_2)a}{V_\mathrm{rel}^2} ,
}
where $(m^\mathrm{bin}_1, m^\mathrm{bin}_2)= (2\times32~\msun, 2\times8~\msun)$ 
are the binary mass. 
Thus the hyper-runaway ejection rate of early B-type stars 
due to 4-body interaction with $\sim 4$ times more massive stars 
in the core region of a YMC is given by
\eq{ \label{eq:Gamma_binbin_with_number}
\Gamma_{\text{bin-bin}} &\simeq 1.49 \times 10^{-4} \Myr^{-1} 
\left( \frac{r_\mathrm{c}}{0.1 \pc} \right)^{-3} 
\left( \frac{M_\mathrm{cl}}{10^4 ~\msun} \right)^{2} \nonumber \\
&\times 
\left( \frac{V_\mathrm{rel}}{5 \kms} \right)^{-1}
\left( \frac{m^\mathrm{bin}_1+m^\mathrm{bin}_2}{80 ~\msun} \right) 
\left( \frac{a}{30 ~R_\odot} \right) .
}
By taking into account that 
this dynamical channel to eject hyper-runaway stars 
is active for
$\sim 6 \Myr$ 
(which is determined by the main-sequence age of 32-$\msun$ star\footnote{
For more massive stars, the main-sequence age is shorter. 
However, they have larger mass and larger cross section, 
so it does not affect our estimation. 
This is another justification for 
treating massive stars with $M=(20{\text -}60)~\msun$ 
as if they are 32-$\msun$ stars. 
} as well as the timescale during which the core density of a YMC is high---see Section \ref{sec:IMF}), 
each YMC ejects $(\Gamma_{\text{bin-bin}}\times 6\Myr)$ massive hyper-runaway stars. 
In order to estimate the total number of massive hyper-runaway stars, 
we need to integrate  
$(\Gamma_{\text{bin-bin}}\times 6\Myr \times 
\mathrm{d}N_\mathrm{cl}/\mathrm{d}M_\mathrm{cl})$ over $M_\mathrm{cl}$ 
at $10^3 ~\msun \leq M_\mathrm{cl} \leq 4\times 10^{5}~\msun$. 
(The cluster mass function is given in equation (\ref{eq:dN_dM}).) 
By doing this integration, 
we estimate that the number of massive hyper-runaway stars 
ejected from the stellar disk in the last $30\Myr$ is 
\eq{\label{eq:N_all_binbin}
N_{\text{bin-bin}} &\simeq 0.943 
\left( \frac{r_\mathrm{c}}{0.1 \pc} \right)^{-3} \nonumber \\
&\times 
\left( \frac{V_\mathrm{rel}}{5 \kms} \right)^{-1}
\left( \frac{m^\mathrm{bin}_1+m^\mathrm{bin}_2}{80 ~\msun} \right) 
\left( \frac{a}{30 ~R_\odot} \right) .
}
Here, we implicitly assume that the core radius $r_\mathrm{c}$ is 
common for all YMCs. 
Since $\Gamma_{\text{bin-bin}}$ scales as $M_\mathrm{cl}^2$, 
a small number of very massive YMCs have 
some impact on $N_{\text{bin-bin}}$. 
For example, 
if the integration range is $10^4 ~\msun \leq M_\mathrm{cl} \leq 4\times 10^{5}~\msun$ 
and $10^5 ~\msun \leq M_\mathrm{cl} \leq 4\times 10^{5}~\msun$, 
the representative number of hyper-runaway ejections 
(which is 0.943 in equation (\ref{eq:N_all_binbin}))
are $N_{\text{bin-bin}} \simeq 0.895$ and $0.517$, respectively. 
This is remarkable given that 
we expect only $\simeq 2$ YMCs more massive than $10^5 ~\msun$
in the last $30\Myr$ (see Section \ref{sec:clusterMF}). 
This also means that 
our estimation on $N_{\text{bin-bin}}$ will be affected by 
Poisson fluctuations on the number of very massive YMCs. 
In any case, our simple analysis suggests that 
the ejection of \hvs, an 8-$\msun$ massive hyper-runaway star with
$V_\mathrm{ej}^\mathrm{int} \sim 600 \kms$, 
can be marginally explained if we consider binary-binary interaction 
at the high-density core region of YMCs.

\subsection{Ejection frequency by an IMBH}
\label{sec:gamma_IMBH}

If a YMC harbors an IMBH, 
it can eject massive hyper-runaway stars through the Hills mechanism
\citep{Hills1988,YuTremaine2002,bromley_etal_06,Gvaramadze2008,Brown2015ARAA}. 
When a binary system with the semi-major axis $a$ 
consisting of stars with mass $(m_1, m_2)$ 
approaches near the tidal breakup radius of 
\eq{
r_t = \left( \frac{3 M_\mathrm{BH}}{m_1+m_2} \right)^{1/3} a,
}
the binary is disrupted with some probability. 
If one of the stars, say the primary star ($m_1$), 
is captured by the IMBH, 
the secondary star ($m_2$) 
is ejected with a large velocity 
due to the energy conservation. 
Numerical experiments (e.g., \citealt{bromley_etal_06})
suggest that 
the intrinsic ejection velocity of a star with mass $m_2$ is 
given by 
\eq{
V_\mathrm{ej}^\mathrm{int} 
&\simeq 521 \kms 
\left( \frac{M_\mathrm{BH}}{100 ~\msun} \right)^{1/6}
\left( \frac{30 ~R_\odot }{a} \right)^{1/2} \nonumber \\
&\times 
\left( \frac{m_1+m_2}{16~\msun} \right)^{1/3}
\left( \frac{2 m_1}{m_1+m_2} \right)^{1/2}
f_R . \label{eq:Vej_Hills}
}
Here $f_R(D) \leq 1$ is an empirical factor of order unity 
that depends on a dimensionless variable
\eq{\label{eq:D}
D = \frac{r_\mathrm{peri}}{a}
\left(\frac{10^6 ~\msun}{M_\mathrm{BH}} \frac{(m_1+m_2)}{2 ~\msun} \right)^{1/3} , 
}
where $r_\mathrm{peri}$ is the pericentric distance of the orbit of the binary 
with respect to the IMBH \citep{Hills1988}. 
The function $f_R(D)$ shows a non-monotonic, unimodal shape, 
and its shape can be inferred from figure 3 of \cite{Hills1988}. 
The approximate expression for $f_R$ is given by 
\eq{\label{eq:fR}
f_R(D) = 
\sum_{k=0}^{5} c_k D^k \;\;(\text{for } 0 \leq D \leq 175) 
}
with $(c_0,c_1,c_2,c_3,c_4,c_5)=(0.774, 0.0204, -6.23\times10^{-4}, 
7.62\times10^{-6}, -4.24\times10^{-8}, 8.62\times10^{-11})$
\citep{bromley_etal_06}. 
The factor $D$ is also related to the 
empirical ejection probability of 
\eq{\label{eq:Pej}
P_\mathrm{ej} = 
\left\{
\begin{array}{ll}
1- D/175 & \;(\text{for } 0 \leq D \leq 175), \\
0        & \;(\text{for } 175<D)
\end{array}
\right.
}
(see figure 1 of \citealt{Hills1988}).

Let us consider the rate of 
hyper-runaway ejection with $V_\mathrm{ej}^\mathrm{int} \geq 600 \kms$ 
of a star with $m_2 = 8 ~\msun$. 
Equation (\ref{eq:Vej_Hills}) implies that  
$V_\mathrm{ej}^\mathrm{int}$ is smaller for larger $a$. 
Given equation (\ref{eq:Vej_Hills}) and the fact that $f_R \leq 1$,  
the condition of $V_\mathrm{ej}^\mathrm{int} \geq 600 \kms$ 
imposes a necessary condition  
\eq{ \label{eq:a_crit}
a \leq a_\mathrm{crit}
\equiv 
&22.6 ~R_\odot \left( \frac{M_\mathrm{BH}}{10^2 ~\msun} \right)^{1/3} \nonumber\\
&\times 
\left( \frac{(m_1+m_2)}{16 ~\msun} \right)^{2/3} 
\left( \frac{2 m_1}{m_1+m_2} \right) . 
}
If $a \leq a_\mathrm{crit}$ is satisfied, 
there exists a certain range of $D$ and therefore $r_\mathrm{peri}$ 
such that $V_\mathrm{ej}^\mathrm{int} \geq 600 \kms$. 
For a given $(M_\mathrm{BH}, m_1, m_2, a)$, 
the allowed range of $r_\mathrm{peri}$ for hyper-runaway ejection 
can be numerically evaluated by using equation (\ref{eq:fR}).
Due to the unimodal shape of $f_R$, 
the allowed range of $r_\mathrm{peri}$ 
can be expressed as\footnote{
If $a$ is exactly equal to $a_\mathrm{crit}$ or slightly smaller than it, 
then the allowed range of $r_\mathrm{peri}$ is very narrow, 
which results in very low probability of hyper-runaway ejection. 
} 
\eq{ \label{eq:rperi_range}
r_\mathrm{peri,min} \leq r_\mathrm{peri} \leq r_\mathrm{peri,max}. 
}
Roughly speaking, the value of $r_\mathrm{peri,max}$ can be regarded as 
the maximum impact parameter for hyper-runaway ejection. 
Since $f_R(D)$ is a non-monotonic function, 
smaller $D$ (or smaller impact parameter) 
does not necessarily result in larger ejection velocity
(see figure 3 of \citealt{Hills1988}). 
Thus, $r_\mathrm{peri,min}$ can have a non-zero value. 
The effective cross section for interaction between the 
binaries with mass $(m_1, m_2)$ and the IMBH 
that results in a hyper-runaway ejection 
is given by 
\eq{\label{eq:sigma_BHbin}
\sigma_\mathrm{BH,bin} \simeq 
\frac{2\pi G (M_\mathrm{BH} + m_1 + m_2) r_\mathrm{peri,max} f_\mathrm{corr}}
{V_\mathrm{rel}^2} . 
}
Here $0 \leq f_\mathrm{corr} \leq 1$ is a correction factor given by 
\eq{
f_\mathrm{corr} = \frac{1}{r_\mathrm{peri,max}} 
\int_{r_\mathrm{peri,min}}^{r_\mathrm{peri,max}} 
\mathrm{d}r \; P_\mathrm{ej}
}
that takes into account 
the ejection probability as a function of $r_\mathrm{peri}$ 
(see equation (\ref{eq:Pej})). 
(If $P_\mathrm{ej}=1$ and $r_\mathrm{peri,min}=0$, 
we have $f_\mathrm{corr} =1$ 
and thus equation (\ref{eq:sigma_BHbin}) 
looks similar to equation (\ref{eq:sigma_binbin}).) 
The collision rate between a given IMBH and the surrounding binaries is  
\eq{
\Gamma_{\text{each-IMBH}} = 
\left( \frac{4\pi}{3} r_\mathrm{c}^3 \right)^{-1} 
N^\mathrm{bin} \sigma_\mathrm{BH,bin} V_\mathrm{rel},
}
where $N^\mathrm{bin}$ is the number of the binaries near the IMBH.

Now let us evaluate $\Gamma_{\text{each-IMBH}}$
for representative environments. 
Here we fix $m_2=8~\msun$, 
$a=30 ~R_\odot$,  
and $V_\mathrm{rel}=5 \kms$. 
We assume that the core radii of YMCs are identical. 
Also, we assume that those YMCs with $M_\mathrm{cl} > 10^4 ~\msun$ 
harbor a $10^2{\text-}\msun$ IMBH, and those with $M_\mathrm{cl} > 10^5 ~\msun$ 
harbor a $10^3{\text-}\msun$ IMBH.

\paragraph{(Case 1)}
If $(M_\mathrm{BH}, m_1) = (10^2~\msun, 8 ~\msun)$, 
the hyper-runaway ejection does not happen since 
$a=30 ~R_\odot \nleq a_\mathrm{crit}=22.6  ~R_\odot$.

\paragraph{(Case 2)}
If $(M_\mathrm{BH}, m_1) = (10^2~\msun, 32 ~\msun)$, 
the allowed range of $r_\mathrm{peri}$ 
(see equation (\ref{eq:rperi_range}))
is $0 \leq r_\mathrm{peri} \leq 72.3 ~R_\odot$ 
and $f_\mathrm{corr} = 0.597$. 
We assume that 
the number of stars with $32 ~\msun$
is given by $N^\mathrm{*}_{20{\text-}60}$ and assume the binary fraction of $80\%$. 
For simplicity, 
we also assume that the binary companion of a 32-$\msun$ star 
is always a 8-$\msun$ star. 
The number of binaries with mass 
$(m_1, m_2)$ in the core of a YMC is given by 
$N^\mathrm{bin} = 0.8 \times N^\mathrm{*}_{20{\text -}60} 
= 1.85 \times 10^{-3} (M_\mathrm{cl}/\msun)$. 
In this case, we obtain 
\eq{
&\Gamma^{(2)}_{\text {each-IMBH}} \simeq 3.32 \times 10^{-3} \Myr^{-1} 
\left( \frac{r_\mathrm{c}}{0.1 \pc} \right)^{-3} 
\left( \frac{M_\mathrm{cl}}{10^4 ~\msun} \right) \nonumber \\
&\times 
\left( \frac{V_\mathrm{rel}}{5 \kms} \right)^{-1}
. %\left( \frac{a}{30 ~R_\odot} \right) 
}
Since the main-sequence lifetime of a 32-$\msun$ star is $\sim 6 \Myr$, 
a 100-$\msun$ IMBH in a YMC ejects 
$N^{(2)}_{\text {each-IMBH}}  = (\Gamma^{(2)}_{\text {each-IMBH}}\times 6 \Myr)$
massive hyper-runaway stars during this $6\Myr$-period. 
If YMCs have an identical core radius $r_\mathrm{c}$, 
then the total number of massive hyper-runaway stars from the stellar disk 
is proportional to the total mass of the YMCs that harbor an IMBH. 
Given that $31.5 \%$ of $M_\mathrm{tot}$ is embedded in 
YMCs with $M_\mathrm{cl}> 10^4 ~\msun$ 
in our cluster mass function in equation (\ref{eq:dN_dM}), 
we estimate that 
\eq{
N^{(2)}_{\text {all-IMBH}} 
\sim 2.96
\left( \frac{r_\mathrm{c}}{0.1 \pc} \right)^{-3} 
\left( \frac{V_\mathrm{rel}}{5 \kms} \right)^{-1}
%\left( \frac{a}{30 ~R_\odot} \right) 
}
massive hyper-runaway stars are ejected from the stellar disk 
in the last $30\Myr$.

\paragraph{(Case 3)}
If $(M_\mathrm{BH}, m_1) = (10^3~\msun, 8 ~\msun)$, 
the allowed range of $r_\mathrm{peri}$ 
is $0.84 ~R_\odot \leq r_\mathrm{peri} \leq 174 ~R_\odot$ 
and $f_\mathrm{corr}=0.664$. 
We assume that the number of stars with $\sim 8 ~\msun$
is given by $N^{*}_{7{\text-}11}$ and 
that the binary fraction is $80\%$. 
For simplicity, 
we also assume that these binaries are equal-mass binaries. 
The number of binaries with mass 
$(8~\msun, 8~\msun)$ in the core of a YMC is given by 
$N^\mathrm{bin} = 0.8 \times 0.5 \times N^\mathrm{*}_{7{\text -}11} 
= 2.26 \times 10^{-3} (M_\mathrm{cl}/\msun)$. 
In this case, we obtain 
\eq{
&\Gamma^{(3)}_{\text {each-IMBH}}  
\simeq 7.86 \times 10^{-2} \Myr^{-1} 
\left( \frac{r_\mathrm{c}}{0.1 \pc} \right)^{-3} 
\left( \frac{M_\mathrm{cl}}{10^4 ~\msun} \right) \nonumber \\
&\times 
\left( \frac{V_\mathrm{rel}}{5 \kms} \right)^{-1}
%\left( \frac{a}{30 ~R_\odot} \right) .
}
Although the main-sequence age of an 8-$\msun$ star is $\sim 30\Myr$, 
we assume that this ejection can last for $6\Myr$, 
by taking into account that the core of YMCs begin to expand 
and the density decreases in the early phase of YMC formation 
\citep{PortegiesZwart2010ARAA}. 
Then a $10^3{\text -}\msun$ IMBH in a YMC ejects 
$N^{(3)}_{\text {each-IMBH}} = (\Gamma^{(3)}_{\text {each-IMBH}} \times 10 \Myr)$
massive hyper-runaway stars during this $6\Myr$-period. 
By assuming that those clusters with $M_\mathrm{cl}>10^5 ~\msun$ 
(that weigh 6.66\% of $M_\mathrm{tot}$) 
harbor a $10^3$-$\msun$ IMBH, we estimate that 
\eq{
N^{(3)}_{\text {all-IMBH}} 
\sim 14.8 
\left( \frac{r_\mathrm{c}}{0.1 \pc} \right)^{-3} 
\left( \frac{V_\mathrm{rel}}{5 \kms} \right)^{-1}
%\left( \frac{a}{30 ~R_\odot} \right) 
}
massive hyper-runaway stars are ejected from the stellar disk 
in the last $30\Myr$.

\paragraph{(Case 4)}
If $(M_\mathrm{BH}, m_1) = (10^3 ~\msun, 32 ~\msun)$, 
then $0 \leq r_\mathrm{peri} \leq 193.4 ~R_\odot$ is allowed 
and $f_\mathrm{corr}=0.500$. 
As in Case 2, we adopt a value of 
$N^\mathrm{bin} = 0.8 \times N^\mathrm{*}_{20{\text -}60} 
= 1.85 \times 10^{-3} (M_\mathrm{cl}/\msun)$. 
In this case, we obtain 
\eq{
&\Gamma^{(4)}_{\text {each-IMBH}} 
\simeq 5.51 \times 10^{-2} \Myr^{-1} 
\left( \frac{r_\mathrm{c}}{0.1 \pc} \right)^{-3} 
\left( \frac{M_\mathrm{cl}}{10^4 ~\msun} \right) \nonumber \\
&\times 
\left( \frac{V_\mathrm{rel}}{5 \kms} \right)^{-1}
. %\left( \frac{a}{30 ~R_\odot} \right) .
}
Since the main-sequence lifetime of a 32-$\msun$ star is $\sim 6 \Myr$, 
a $10^3{\text -}\msun$ IMBH in a YMC ejects 
$N^{(4)}_{\text {each-IMBH}} = (\Gamma^{(4)}_{\text {each-IMBH}} \times 6 \Myr)$
massive hyper-runaway stars during this $6\Myr$-period. 
As in Case 2, we estimate that 
\eq{
N^{(4)}_{\text {all-IMBH}}  
\sim 10.4 
\left( \frac{r_\mathrm{c}}{0.1 \pc} \right)^{-3} 
\left( \frac{V_\mathrm{rel}}{5 \kms} \right)^{-1}
%\left( \frac{a}{30 ~R_\odot} \right) 
}
massive hyper-runaway stars are ejected from the stellar disk 
in the last $30\Myr$.

\subsection{Ejection frequency by a very massive star (VMS)}
\label{sec:gamma_VMS}

If a YMC harbors a VMS, 
it can eject massive hyper-runaway stars through the Hills mechanism, 
just as an IMBH does. 
Here we explain two differences 
between the hyper-runaway ejection 
by an IMBH and that by a VMS. 

First, 
in the VMS ejection, 
the stellar binary has to have $\rperi$ 
that is larger than the radius of the VMS, $R_\mathrm{VMS}$; 
otherwise the binary may merge with the VMS \citep{Gvaramadze2009}. 
In the following discussion, 
we assume that a 100-$\msun$ star has an radius of 
$18~R_\odot$ \citep{Ishii1999}. 
We note that VMSs with $M \gtrsim 133 ~\msun$ 
\citep{Ishii1999,Yungelson2008} 
are predicted to show a core-halo configuration 
\citep{Kato1985} 
such that the mass contribution from the 
stellar outer diffuse layer is negligible. 
Following \cite{Gvaramadze2009}, 
we assume that the mass of a $10^3 {\text -}\msun$ VMS 
is effectively contained within a radius of $40 ~R_\odot$.

Second, a VMS,  
unlike an IMBH, is not expected to survive for a long time. 
The finite age of a VMS determines the 
duration of time when a VMS 
ejects hyper-runaway stars. 
When a VMS forms through continuous mergers 
of massive stars, it can survive longer---due to the rejuvenation---than an isolated main-sequence star with the same mass. 
Since the detailed modeling of VMSs is still not satisfactory 
due to the uncertainty in the mass loss rate of VMSs \citep{Vink2015} 
and in the rejuvenation process \citep{Schneider2016},
in the following,
we simply assume that 
the lifetime of a VMS of any mass is $5 \Myr$, 
motivated by the results in \cite{PortegiesZwart1999}. 
For simplicity, we also assume that 
a VMS is formed instantaneously just after the formation of a YMC
and its mass is constant as a function of time. 
This assumption means that in our toy VMS model 
the stellar mass loss is compensated by the growth of the mass as a result of mergers.

In the following, 
we consider 
Cases 2, 3, and 4 
as in Section \ref{sec:gamma_IMBH}. 
We do not consider Case 1,
since obviously it results in no massive hyper-runaway stars. 
In addition to the above-mentioned assumptions, 
we assume that those YMCs with $M_\mathrm{cl} > 10^4 ~\msun$ 
harbor a $10^2{\text-}\msun$ VMS, 
and those with $M_\mathrm{cl} > 10^5 ~\msun$ 
harbor a $10^3{\text-}\msun$ VMS, 
as in Section \ref{sec:gamma_IMBH}.

\paragraph{(Case 2)}

If $(M_\mathrm{VMS}, m_1) = (10^2~\msun, 32 ~\msun)$, 
the allowed range of $\rperi$ is altered to 
$(R_\mathrm{VMS}=)18 ~R_\odot 
\leq r_\mathrm{peri} \leq 72.3 ~R_\odot$ 
and we have $f_\mathrm{corr}=0.373$. 
We obtain
$\Gamma^{(2)}_{\text {each-VMS}} 
= 0.625 \Gamma^{(2)}_{\text {each-IMBH}}$. 
If we assume that a $10^2$-$\msun$ VMS in a YMC survives for $5 \Myr$, 
it ejects 
$N^{(2)}_{\text {each-VMS}} = 0.521 N^{(2)}_{\text {each-IMBH}} $
massive hyper-runaway stars during this period. 
By repeating the same arguments as in Section \ref{sec:gamma_IMBH}, 
we estimate that 
\eq{
N^{(2)}_{\text {all-VMS}}  
\sim 1.54
\left( \frac{r_\mathrm{c}}{0.1 \pc} \right)^{-3} 
\left( \frac{V_\mathrm{rel}}{5 \kms} \right)^{-1}
%\left( \frac{a}{30 ~R_\odot} \right) 
}
massive hyper-runaway stars are ejected from the stellar disk 
in the last $30\Myr$.

\paragraph{(Case 3)}

If $(M_\mathrm{VMS}, m_1) = (10^3~\msun, 8 ~\msun)$, 
the allowed range of $\rperi$ is altered to 
$(R_\mathrm{VMS}=)40 ~R_\odot 
\leq r_\mathrm{peri} \leq 174.0 ~R_\odot$ 
and we have $f_\mathrm{corr}=0.456$. 
We obtain 
$\Gamma^{(3)}_{\text {each-VMS}} 
= 0.687 \Gamma^{(3)}_{\text {each-IMBH}}$. 
If we assume that a $10^3$-$\msun$ VMS in a YMC survives for $5 \Myr$, 
it ejects 
$N^{(3)}_{\text {each-VMS}} = 0.573 N^{(3)}_{\text {each-IMBH}} $
massive hyper-runaway stars during this period. 
By repeating the same arguments, 
we estimate that 
\eq{
N^{(3)}_{\text {all-VMS}} 
\sim 8.48
\left( \frac{r_\mathrm{c}}{0.1 \pc} \right)^{-3} 
\left( \frac{V_\mathrm{rel}}{5 \kms} \right)^{-1}
%\left( \frac{a}{30 ~R_\odot} \right) 
}
massive hyper-runaway stars are ejected from the stellar disk 
in the last $30\Myr$.

\paragraph{(Case 4)}

If $(M_\mathrm{VMS}, m_1) = (10^3~\msun, 32 ~\msun)$, 
the allowed range of $\rperi$ is altered to 
$(R_\mathrm{VMS}=)40 ~R_\odot 
\leq r_\mathrm{peri} \leq 193.4 ~R_\odot$ 
and we have $f_\mathrm{corr}=0.315$. 
We obtain
$\Gamma^{(4)}_{\text {each-VMS}} 
= 0.629 \Gamma^{(4)}_{\text {each-IMBH}}$. 
If we assume that a $10^3$-$\msun$ VMS in a YMC survives for $5 \Myr$, 
it ejects 
$N^{(4)}_{\text {each-VMS}} = 0.524 N^{(4)}_{\text {each-IMBH}} $
massive hyper-runaway stars during this period. 
By repeating the same arguments, 
we estimate that 
\eq{
N^{(4)}_{\text {all-VMS}} 
\sim 5.45
\left( \frac{r_\mathrm{c}}{0.1 \pc} \right)^{-3} 
\left( \frac{V_\mathrm{rel}}{5 \kms} \right)^{-1}
%\left( \frac{a}{30 ~R_\odot} \right) 
}
massive hyper-runaway stars are ejected from the stellar disk 
in the last $30\Myr$.

\subsection{Summary and discussion 
on the massive hyper-runaway star ejection frequency} 
\label{sec:summary_ejection}

In Sections 
\ref{sec:gamma_massive}, 
\ref{sec:gamma_IMBH}, and 
\ref{sec:gamma_VMS}, 
we have explored the ejection frequency of 
massive ($\gtrsim 8 \msun$) hyper-runaway stars with 
$V_\mathrm{ej}^\mathrm{int} \gtrsim 600 \kms$ 
(comparable to the ejection velocity of \hvs). 
Since the ejection rate is proportional to 
the number density of massive binaries (for IMBH and VMS ejection) or 
the number density squared (for binary-binary interaction),  
massive hyper-runaway ejection 
requires a high density. 
Also, since all of these mechanisms are involved with massive objects, 
hyper-runaway ejection must occur  
in young star-forming regions. 
These requirements seem to indicate that 
massive hyper-runaway ejection happens only in YMCs.

We estimate that the number of massive hyper-runaway stars
ejected in the last $30 \Myr$ from the stellar disk of the Milky Way is 
$\sim 1$ for binary-binary encounter of ordinary stars;
$\sim {\text {(3-15)}}$ for ejection by an IMBH; and 
$\sim {\text {(2-8)}}$ for ejection by a VMS. 
Since \hvs\ is the only well-confirmed example of a massive hyper-runaway star with $V_\mathrm{ej}^\mathrm{int} \gtrsim 600 \kms$ 
that was ejected in the last $\sim 30\Myr$, 
it is interesting that 
all of the above-mentioned dynamical channels 
can explain the ejection of \hvs. 
However, we should bear in mind that 
we have made some simplistic assumptions 
in estimating the ejection frequency. 
For example, we can easily change the ejection frequency 
by a factor of a few with some fine-tuning of the parameters used.
Some obvious sources of uncertainties include 
the influence from the adopted IMF (Section \ref{sec:IMF}) 
and the mass function of the YMCs (Section \ref{sec:clusterMF}).

Although the absolute value of our ejection frequency 
may not be very accurate, 
our results are still useful in understanding the relative efficiency 
in ejecting hyper-runaway stars. 
Here we compare the ejection by an IMBH and VMS. 
Our results suggest that 
a 100-$\msun$ IMBH is $\sim 2$ times more efficient 
in ejecting hyper-runaway stars than a 100-$\msun$ VMS. 
This difference arises from the fact that 
a very close encounter with an IMBH 
may result in a hyper-runaway star, 
while 
a very close encounter with a VMS will end up with a merger instead. 
Of course, this result alone does not favor IMBH-origin over VMS-origin,  
since 
forming a 100-$\msun$ IMBH is probably more difficult than 
forming a VMS with the same mass. 
Also, we note that no IMBH has been detected in YMCs in the Milky Way, 
while some massive stars in YMCs are regarded as remnants of VMSs.

Our estimate of the ejection frequency 
by binary-binary interactions 
is smaller than those by an IMBH or a VMS. 
However, we regard this ejection channel promising, 
since it requires multiple ordinary massive stars ($\sim 30~\msun$) and 
it does not require any exotic objects or phenomena. 
Our simple estimate suggests that 
the ejection of massive hyper-runaway star 
with $V_\mathrm{ej}^\mathrm{int} \gtrsim 600 \kms$ 
can occur every $\sim 30 \Myr$. 
Since the ejection frequency in this scenario 
is proportional to the square of the number density of massive binaries 
(see equations (\ref{eq:Gamma_binbin_encounter}) and (\ref{eq:Gamma_binbin})), 
the ejection rate can be enhanced by a factor of $X^2$ 
if the number density of massive objects is increased 
by a factor of $X$ by adopting a top-heavy IMF. 
Also, since 
the ejection frequency from a given YMC 
is proportional to $M_\mathrm{cl}^2$ 
(see equation (\ref{eq:Gamma_binbin_with_number})), 
the total number of hyper-runaway stars is increased 
if a larger number of very massive YMCs 
have been formed in the last $30\Myr$. 
Such an over-production of very massive YMCs 
can happen as a result of Poisson fluctuations. 
Indeed, under our assumption of the cluster mass function, 
the number of massive clusters with mass $10^5 ~\msun < M_\mathrm{cl} < 10^{5.6}~\msun$ 
formed in the last $30\Myr$ is expected to be $2$ (Section \ref{sec:clusterMF}), 
so the Poisson fluctuation may result in 70\% ($=\sqrt{2}/2)$ 
overproduction or underproduction of YMCs with
$10^5 ~\msun < M_\mathrm{cl} < 10^{5.6}~\msun$. 
This results in 40\% ($=(\sqrt{2}/2) \times 0.517 / 0.943$) 
overproduction or underproduction of 
massive hyper-runaway stars. 
In any case, 
our calculations encourage more detailed simulations of 
4-body interaction of ordinary massive stars in YMCs 
dedicated to study massive hyper-runaway stars 
\citep{Leonard1991,Fujii2011,Perets2012}.

If our estimate of the ejection frequency 
is correct to a factor of a few, 
IMBHs or VMSs in the Milky Way may have ejected 
at least a few more massive hyper-runaway stars 
with $V_\mathrm{ej}^\mathrm{int} \gtrsim 600 \kms$. 
Also, there is some room for the 
4-body interaction of ordinary massive stars 
to have ejected a few more massive hyper-runaway stars. 
If more massive hyper-runaway stars are discovered or confirmed in the future, 
then it will further motivate us to study the 
ejection mechanism of massive hyper-runaway stars.

\section{Discussion and conclusion}
\label{sec:discussion_conclusion}

We obtained high-resolution spectra of \hvs\ with 
MIKE spectrograph mounted on Magellan Telescope 
and confirmed that this star is an 8.3-$\msun$ B-type subgiant star 
with super-Solar metallicity 
([Si/H]$=0.60\pm0.06$, [Mg/H]$=0.33\pm0.10$; Table \ref{table:abundances}). 
By using the 
proper motion from \Gaia\ DR2 
and the spectroscopic distance and line-of-sight velocity 
(Table \ref{table:stellar_parameters}), 
we reconstruct the orbit of this star. 
The orbital analysis 
suggests that 
this star was ejected from the inner stellar disk of the Milky Way 
$(R_\mathrm{ej} = 2.9_{-0.9}^{+2.5} \kpc)$
with the intrinsic ejection velocity 
(corrected for the disk streaming motion) 
$V_\mathrm{ej}^\mathrm{int} = 568^{+19}_{-17} \kms$ 
(Figures \ref{fig:triangle_Vej_Rej_flightTime_dist} and \ref{fig:xy}). 
The ejection happened $\sim 33 \Myr$ ago, 
probably just after the formation of this star in the natal star cluster. 
{\textcolor{black}{
The chemical abundance pattern of \hvs\ 
is similar to that of inner-disk young massive stars 
(see Table \ref{table:abundances} and Section \ref{sec:chemistry_disk}), 
which supports the disk-origin of this star. 
}}

The large intrinsic ejection velocity of \hvs\ 
rules out the possibility that this star was ejected by 
the supernova explosion of the binary companion. 
Rather, this star was probably ejected by 
a few-body dynamical interaction with more massive objects 
in a high-density environment. 
Such an extreme environment may be attained  
in the core region of a YMC 
with mass of $\gtrsim 10^4 ~\msun$.

Based on the flight time and the ejection location of \hvs, 
we argue that its ejection agent 
(the other object that took part in this ejection) 
or the natal star cluster  
must be currently located in the inner disk, 
probably near the Norma spiral arm 
(Figure \ref{fig:triangle_distAgent_ellAgent}). 
If the natal star cluster of \hvs\ is as massive as $\sim 10^4 ~\msun$, 
we expect that the cluster survives until today. 
The current locations and ages of the known YMCs 
as well as the estimated current location of the natal star cluster of \hvs\ 
indicate that 
none of the known YMCs can be the natal cluster of \hvs\  
and therefore it may imply the existence of an undiscovered YMC.

\hvs\ is the first well-confirmed 
early B-type massive ($\gtrsim 8 ~\msun$) 
hyper-runaway star with 
$V_\mathrm{ej}^\mathrm{int} \gtrsim 600 \kms$ 
ejected from the inner Galactic disk. 
This fact 
suggests that the ejection frequency of such massive hyper-runaway stars 
from the entire stellar disk of the Milky Way is 
$\gtrsim 1$ per $30 \Myr$. 
We argue that 
this rough estimate of the ejection frequency can be explained 
by a few-body interaction associated with either 
an IMBH ($\gtrsim 100 ~\msun$), a very massive star ($\gtrsim 100 ~\msun$), 
or multiple ordinary massive stars ($\gtrsim 30 ~\msun$) 
(Sections \ref{sec:mechanism} and \ref{sec:rate}). 
If more massive hyper-runaway stars are discovered  
in the future, it will help us understand 
the origin of massive hyper-runaway stars 
and the dynamics of the extreme environment in which massive stars form.

%\phantom{.}
\acknowledgments

We thank members of the stellar halos group at the University of Michigan for stimulating discussion. 
We thank Mark Reynolds for discussion on Galactic X-ray sources. 
We thank Mark Gieles for providing us with the coordinate information of YMCs. 
We thank the staff and workers at the 6.5m Magellan Telescopes at the Las Campanas Observatory, Chile, for their labor.
M.V. and K.H. are supported by NASA-ATP award NNX15AK79G.
I.U.R. acknowledges support from 
grants PHY~14-30152 (Physics Frontier Center/JINA-CEE), 
AST~16-13536, and AST~18-15403 
awarded by the U.S.\ National Science Foundation (NSF).~
This research was started at the NYC {\it Gaia} DR2 Workshop at the Center for Computational Astrophysics of the Flatiron Institute in 2018 April. 
This work has made use of data from the European Space Agency (ESA)
mission {\it Gaia} 
(\url{http://www.cosmos.esa.int/gaia}), 
processed by the {\it Gaia} Data Processing and Analysis Consortium (DPAC,
\url{http://www.cosmos.esa.int/web/gaia/dpac/consortium}). 
Funding for the DPAC has been provided by national institutions, in particular
the institutions participating in the {\it Gaia} Multilateral Agreement. 
%LAMOST
Guoshoujing Telescope (the Large Sky Area Multi-Object Fiber Spectroscopic Telescope LAMOST) is a National Major Scientific Project built by the Chinese Academy of Sciences. Funding for the project has been provided by the National Development and Reform Commission. LAMOST is operated and managed by the National Astronomical Observatories, Chinese Academy of Sciences.
%other
This research has made use of NASA's Astrophysics Data System Bibliographic Services; the arXiv pre-print server operated by Cornell University; the SIMBAD and VizieR databases hosted by the Strasbourg Astronomical Data Center.

\facility{Gaia, Magellan}

\software{
Agama \citep{Vasiliev2018},\;
corner.py \citep{ForemanMackey2016}, 
matplotlib \citep{hunter07},
numpy \citep{vanderwalt11},
scipy \citep{jones01}}

\bibliographystyle{aasjournal}
\bibliography{mybibtexfile}
% To add a second bibtex file, separate the name (no suffix) 
% by a comma with no space in the command above.

\appendix

\section{A note on the quality of the astrometric solution} \label{sec:Gaia_quality}

{\textcolor{black}{
When using the \Gaia\ data, we need to be careful about the 
quality of the astrometric solution. 
To assess whether the astrometric data for \hvs\ are reliable, 
we checked some statistical indicators in \Gaia\ DR2 for this star. 
}}

{\textcolor{black}{
First, 
we checked the 
criteria (i)-(iv) in Section 4 of \cite{Marchetti2018}, 
which are based on the \Gaia\ documentation as of April 2018 
(when \Gaia\ DR2 was released):
\begin{itemize}
\item[] (i)   ASTROMETRIC\_GOF\_AL $< 3$;
\item[] (ii)  ASTROMETRIC\_EXCESS\_NOISE\_SIG $\leq 2$;
\item[] (iii) $-0.23 \leq$  MEAN\_VARPI\_FACTOR\_AL $\leq 0.32$; 
\item[] (iv) VISIBILITY\_PERIODS\_USED $> 8$;
\end{itemize}
For \hvs, we have 
ASTROMETRIC\_GOF\_AL $=22.03$, 
ASTROMETRIC\_EXCESS\_NOISE\_SIG $=32.46$, 
MEAN\_VARPI\_FACTOR\_AL $=-0.06760$, 
and
VISIBILITY\_PERIODS\_USED $=9$; 
therefore the criteria (i) and (ii) are not satisfied 
but (iii) and (iv) are satisfied. 
}}

{\textcolor{black}{
Second, 
we checked the distributions of 
(i) ASTROMETRIC\_GOF\_AL 
and 
(ii) ASTROMETRIC\_EXCESS\_NOISE\_SIG 
for those stars for which the $G$ magnitude is similar to that of \hvs. 
Given that the reported parallax for \hvs\ is negative, 
we checked the distributions of these quantities for 
stars with positive parallax and negative parallax. 
As a result, we found that 
both ASTROMETRIC\_GOF\_AL and ASTROMETRIC\_EXCESS\_NOISE\_SIG 
are typically much larger than the \Gaia's recommended upper limits 
(3 and 2, respectively) 
when the parallax is negative. 
This indicates that the large values of ASTROMETRIC\_GOF\_AL 
and ASTROMETRIC\_EXCESS\_NOISE\_SIG for \hvs\
may be related to the negative parallax of this star. 
(It is not surprising that 
the goodness-of-fit of an astrometric solution is poor 
if the reported parallax is negative.) 
}}

{\textcolor{black}{
Lastly, 
we checked the quality flag that was more recently proposed. 
According to 
\cite{Lindegren2018RUWE}
(\Gaia\ DPAC document published in August 2018), 
a quantity called 
`renormalized unit weight error' ($\textrm{RUWE}$) 
may be a better way of assessing the quality of the astrometric solution. 
Here, $\textrm{RUWE}$ is given by 
\eq{
\textrm{RUWE} = \frac{1}{
u_0(G, G_\mathrm{BP} - G_\mathrm{RP})
}
\sqrt{
\frac
{\textrm{ASTROMETRIC\_CHI2\_AL}}
{\textrm{ASTROMETRIC\_N\_GOOD\_OBS\_AL}-5}, 
}
}
and $u_0$ is a normalization factor. 
From the available data of
{ASTROMETRIC\_CHI2\_AL} $=916.65$, 
{ASTROMETRIC\_N\_GOOD\_OBS\_AL} $=141$, 
$G = 13.06$, 
$(G_\mathrm{BP} - G_\mathrm{RP}) = -0.2316$, and 
$u_0(G, G_\mathrm{BP} - G_\mathrm{RP}) = 2.11399$ for this star,\footnote{
The value of $u_0$ is obtained from the interpolation of the look-up table 
available at \url{https://www.cosmos.esa.int/web/gaia/dr2-known-issues}. 
}
we obtain $\textrm{RUWE} =1.228$. 
The value of $\textrm{RUWE}$ is close to $1.0$ 
and smaller than $1.4$, 
which is expected for a well-behaved astrometric solution \citep{Lindegren2018RUWE}. 
Thus, the astrometric solution for \hvs\ is probably well-behaved 
given its magnitude and color. 
}}

{\textcolor{black}{
Based on these assessments (especially the asessment of $\textrm{RUWE}$), 
we think that the use of \Gaia\ proper motion for this star is justified. 
The rather large values of the above-mentioned flags (i) and (ii) 
are probably related to the negative parallax, 
but we note that the \Gaia\ parallax is not used in our analysis. 
We will revisit the orbital analysis of \hvs\ with future data releases, 
for which the astrometric data are expected to be improved. 
}}

\end{document}